\documentclass[10pt]{article}
\usepackage{citesort}
\usepackage{amsmath}
\usepackage{amssymb}
\usepackage{latexsym}
\usepackage{amssymb}
\usepackage{epsfig}
\usepackage{natbib}
\usepackage{color}
\usepackage{enumerate}
\usepackage{epstopdf}
\usepackage{xcolor}
\usepackage{hyperref} 
\hypersetup{
allcolors       = {cyan},
linkbordercolor = {white},
linkcolor       = {blue},
colorlinks      = true,
}
\newcommand*{\doi}[1]{\href{http://dx.doi.org/#1}{doi: #1}}
\def \ii{{\mathrm{i}}}
\def \TT{{\mathrm{T}}}
\def \TL{{\mathrm{L}}}
\def \TP{{\mathrm{P}}}

\def \d{{\mathrm{d}}}

\def \R{{\mathbb{R}}}

\def \pd{\partial}

\def \Bsigma{\boldsymbol{\sigma}}
\def \Bbeta{{\boldsymbol{\beta}}}

\def \BI{\boldsymbol{I}}

\def \Bs{\boldsymbol{s}}

\def \Bb{{\boldsymbol{b}}}

\def \Bu{{\boldsymbol{u}}}

\def \Bv{{\boldsymbol{v}}}
\def \BR{{\boldsymbol{R}}}
\def \BV{{\boldsymbol{V}}}

\def \Bbeta{\boldsymbol{\beta}}
\def \Balpha{\boldsymbol{\alpha}}
\def \rr{{\boldsymbol{r}}}

\def \Bu{{\boldsymbol{u}}}
\def \Bv{{\boldsymbol{v}}}

\def \Bp{{\boldsymbol{p}}}
\def \tt{{\bar{t}}}
\renewcommand\Re{\mathop{\text{Re}}}

\def \ii{{\text{i}}}

\def \sign{\mathop{\text{sign}}}

\def \cT{c_{\text{T}}}
\def \cL{c_{\text{L}}}

\newcommand{\bbeta}{\boldsymbol{\beta}}

\def \br{\rr}
\def \bV{\BV}
\def \bhr{\widehat{\rr}}
\def \hr{\widehat{r}}
\newcommand{\hn}{\widehat{n}}
\newcommand{\bhn}{\mathbf{\widehat{n}}}

\begin{document}
\title{{\bf
Distributional and regularized radiation fields of
non-uniformly moving straight dislocations,
and elastodynamic Tamm problem
}}
\author{
Markus Lazar~$^\text{a,}$\footnote{
{\it E-mail address:} lazar@fkp.tu-darmstadt.de.}
\
and Yves-Patrick Pellegrini~$^\text{b,}$\footnote{
{\it E-mail address:} yves-patrick.pellegrini@cea.fr.}
\\ \\
${}^\text{a}$
        Heisenberg Research Group,\\
        Department of Physics,\\
        Darmstadt University of Technology,\\
        Hochschulstr. 6,\\
        D-64289 Darmstadt, Germany\\
\\
${}^\text{b}$
CEA, DAM, DIF,\\
F-91297 Arpajon, France
}
\date{\vspace{-5ex}}

\maketitle
\begin{abstract}
This work introduces original explicit solutions for the elastic fields radiated by non-uniformly moving, straight, screw or edge dislocations in an isotropic medium, in the form of time-integral representations in which acceleration-dependent contributions are explicitly separated out. These solutions are obtained by applying an isotropic regularization procedure to distributional expressions of the elastodynamic fields built on the Green tensor of the Navier equation. The obtained regularized field expressions are singularity-free, and depend on the dislocation density rather than on the plastic eigenstrain. They cover non-uniform motion at arbitrary speeds, including faster-than-wave ones. A numerical method of computation is discussed,
that rests on discretizing motion along an arbitrary path in the plane transverse to the dislocation, into a succession of time intervals of constant velocity vector over which time-integrated contributions can be obtained in closed form. As a simple illustration, it is applied to the elastodynamic equivalent of the Tamm problem, where fields induced by a dislocation accelerated from rest beyond the longitudinal wave speed, and thereafter put to rest again, are computed. As expected, the proposed expressions produce Mach cones, the dynamic build-up and decay of which is illustrated by means of full-field calculations.
\\

\end{abstract}

\newpage
\section{Introduction}
Dislocations are linear defects whose motion is responsible for plastic deformation in crystalline materials \citep{HIRT82}. To improve the current understanding of the plastic and elastic fronts \citep{CLIF81} that go along with extreme shock loadings in metals \citep{MEYE09}, \cite{Gurr13} recently proposed to make dynamic simulations of large sets of dislocations mutually coupled by their retarded elastodynamic field. \cite{Gurr14} review the matter and its technical aspects in some detail. This new approach is hoped to provide complementary insights over multi-physics large-scale atomistic simulations of shocks in matter \citep{ZHAK11}. If we leave aside the subsidiary (but physically important) issue of dislocation nucleation, dislocation-dynamics simulations involve two separate but interrelated tasks. First, one needs to compute the field radiated by a dislocation that moves arbitrarily. Second, given the past history of each dislocation, the current dynamic stress field incident on it due to the other ones, and the externally applied stress field (e.g., a shock-induced wavefront), the further motion of the dislocation must be determined by a dynamic mobility law. While some progress has recently been achieved in the latter subproblem ---which involves scarcely explored radiation-reaction effects and dynamic core-width variations \citep{PELL14}--- the focus of the present paper is on the former ---a very classical one.

Indeed, substantial effort has been devoted over decades to obtaining analytical expressions of elastodynamic fields produced by non-uniformly moving singularities such as point loads \citep{STRO70,FREU72,FREU73}, cracks, and dislocations.  Results ranged, e.g., from straightforward applications to linear-elastic and isotropic unbounded media, to systems with interfaces such as half-spaces (Lamb's problem) or layered media \citep{EATW82}; coupled phenomena such as thermoelasticity \citep{BROC97} or anisotropic elastic media \citep{MARK87,Wu}, to mention but a few popular themes. Elastodynamic fields of dislocations have been investigated in a large number of works, among which \citep{ESHE51b,KM64,KM65,Mura,Nabarro,BROC79a,Brock82,Brock83,XM80,XM01,XM01b,Pellegrini10,Lazar2011,Lazar12,Lazar13,Lazar2013}. Early numerical implementations of time-dependent fields radiated by moving sources \citep{NIAZ75,MADA78} were limited to material displacements or velocities. As to stresses, \cite{Gurr14} based their simulations on the fields of \cite{XM81} relative to a subsonic edge dislocation. Nowadays dynamic fields of individual dislocations or cracks are also investigated by atomistic simulations \citep{LISH02,TSUZ09,SPIE09}, or numerical solutions of the wave equation by means of finite-element \citep{ZHAN15}, finite-difference, or boundary-integral schemes \citep{DAYD05}. Hereafter, the analytical approach is privileged so as to produce reference solutions.

Disregarding couplings with other fields such as temperature, one might be tempted to believe that the simplest two-dimensional problem of the non-uniform motion of rectilinear dislocation lines in an unbounded, linear elastic, isotropic medium, leaves very little room for improvements over past analytical works. This is not so, and our present concerns are as follows:
\smallskip

(i) \emph{Subsonic as well as supersonic velocities.} In elastodynamics, from the 70's onwards, the method of choice for analytical solutions has most often been the one of Cagniard improved by de Hoop \citep{AKIR09}, whereby Laplace transforms of the fields are inverted by inspection after a deformation of the integration path of the Laplace variable has been carried out by means of a suitable change of variable (see above-cited references). However, to the best of our knowledge, no such solutions can be employed indifferently for subsonic and supersonic motions, in the sense that the supersonic case need be considered separately in order to get explicit results as, e.g., in \citep{STRO70,FREU72,CALL80,XM01b,HUAN11}. Indeed, carrying out the necessary integrals usually requires determining the wavefront position relatively to the point of observation. To date, the supersonic edge dislocation coupled to both shear and longitudinal waves has not been considered, and existing supersonic analytical solutions for the screw dislocation have not proved usable in full-field calculations, except for the rather different solution obtained within the so-called gauge-field theory of dislocations \citep{Lazar09}, which appeals to gradient elasticity. Thus, one objective of the present work is to provide `automatic' theoretical expressions that do not require wavefront tracking, for both screw and edge dislocations, and can be employed whatever the dislocation velocity. To this aim, we shall employ a method different from the Cagniard--de Hoop one. This is not to disregard the latter but following a different route was found more convenient in view of the remaining points listed here.

(ii) \emph{Distributions and smooth regularized fields.} For a Volterra dislocation in supersonic steady motion, fields are typically concentrated on Dirac measures along infinitely thin lines, to form Mach cones \citep{STRO70,CALL80}. Thus, the solution is essentially of distributional nature, and its proper characterization involves, beside Dirac measures, the use of principal-value and finite-parts prescriptions \citep{PL2014}. Of course, in-depth analytical characterizations of wavefronts singularities can still be extracted out of Laplace-transform integral representations \citep{FREU72,FREU73,CALL80}. However their distributional character implies that the solutions cannot deliver meaningful numbers unless they are regularized by convolution with some source shape function representing a dislocation of finite width. Only by this means can field values in Mach cones be computed. Consequently, another objective is to provide field expressions for an extended dislocation of finite core width (instead of a Volterra one), thus taming all the field singularities that would otherwise be present at wavefronts and at the dislocation location, where Volterra fields blow up. In the work by \cite{Gurr13}, a simple cut-off procedure was employed to get rid of infinities. Evidently, a similar device cannot be used with Dirac measures, which calls for a smoother and more versatile regularization. Various dislocation-regularizing devices have been proposed in the past, some consisting in expanding the Volterra dislocation into a flat Somigliana dislocation \citep{ESHE49,ESHE51b,XM01,XM01b,Pellegrini11}. Such regularizations remove infinities, but leave out field discontinuities on the slip path \citep{ESHE49}. A smoother approach consisting in introducing some non-locality in the field equations has so far only be applied to the time-dependent motion of a screw dislocation. The one to be employed hereafter, introduced in \citep{PL2014}, achieves an isotropic expansion the Volterra dislocation and smoothly regularizes all field singularities for screw and edge dislocations. In this respect, it resembles that introduced in statics by \cite{CAIA06}. However, we believe it better suited to dynamics.

(iii) \emph{Field-theoretic framework.}  The traditional method of solution \citep{XM80} rests on imposing suitable boundary conditions on the dislocation path. It makes little contact with field-theoretic notions of dislocation theory such as plastic strain, or dislocation density and current used in purely numerical methods of solution \citep{Djaka15}. Instead, we wish our analytical results to be rooted on a field-theoretic background. One advantage is that the approach will provide a representation of radiation fields where velocity- and acceleration-dependent contributions are \emph{clearly separated out}, which is most convenient for subsequent numerical implementation. Again to the best of our knowledge, no such representation of the elastodynamic fields has been given so far. However, previous work in that direction can be found in \citep{Lazar2011,Lazar12,Lazar13}.

(iv) \emph{Integrals in closed form.} In \citep{Gurr13} the numerical implementation of the results by \cite{XM81}, where the retarded fields are expressed in terms of an integral over the path abscissa, is not fully explicit. Indeed, this integral is split over path segments, and each segment is integrated over numerically --- a tricky matter, as pointed out by the former authors. By contrast, and dealing with time intervals instead of path segments, the sub-integrals will be expressed hereafter in closed form by means of the key indefinite integrals obtained in \citep{PL2014}.

(v) \emph{Arbitrary paths.} Results will be given in tensor form, with the dislocation velocity as a vector. Thus, they can be applied immediately to arbitrary dislocation paths parametrized by time. Using the time variable as the main parameter is a natural choice, and does not require computing so-called `retarded times'. Although we must leave such applications to further work, this makes it straightforward to investigate radiative losses in various oscillatory motions, e.g., (lattice-induced) periodic oscillations in the direction transverse to the main glide plane during forward motion, which space-based parametrizations such as in the procedure outlined by \cite{Brock83} make harder to achieve.
\smallskip

Accordingly, our work is organized as follows. First, we begin by computing in Section \ref{sec2} general forms for the elastic fields of non-uniformly moving screw and edge dislocations using the theory of distributions, starting from the most general field equations in terms of dislocation densities and currents. Our approach relies on Green's functions [e.g.,~\citet{Barton,Mura}]. In Section \ref{sec2-1}, inhomogeneous Navier equations  for the elastic fields are derived as equations of motion, with source terms expressed in terms of the fields that characterize the dislocation (dislocation density tensor and dislocation current tensor). The Cauchy problem of the Navier equations is then addressed in Section \ref{sec2-2}, where the solutions for the elastic fields are written as the convolution of the retarded elastodynamic Green-function tensor ---interpreted as a distribution--- with the dislocation fields. As a result, the mathematical structure of the latter is partly inherited from the former. Some connections with past works are made in Section \ref{sec:remarks}.
Second, we specialize the obtained field expressions to Volterra dislocations: the fields themselves become distributions. In Section \ref{sec3}, the structure of the Green tensor and of the elastodynamic radiation fields is revealed and analyzed in terms of locally-integrable functions and pseudofunctions (namely, singular distributions that require a `finite part' prescription). The Volterra screw (Section \ref{sec3-1}) and edge (Section \ref{sec3-2}) dislocations are addressed separately for definiteness. The expressions reported are mathematically well-defined, and cover arbitrary speeds including faster-then-wave ones, which is the main difference with classical approaches.
Third, since distributional fields, although mathematically correct, cannot in general produce meaningful numbers unless being applied to test functions, we turn the formalism into one suitable to numerical calculations by means of the isotropic-regularization procedure alluded to above, where the relevant smooth test function represents the dislocation density. The procedure is introduced in Section \ref{sec4-1}, and regularized expressions for the elastic fields are obtained in integral form in Section \ref{sec4-2}, after the regularized Green tensor has been defined. Next, a numerical implementation scheme that involves only closed-form results is proposed in Section~\ref{sec5}, based on the key integrals of \cite{PL2014}. As a first illustration, the particular case of steady motion for the edge dislocation is discussed in detail, with emphasis on faster-than-wave motion. Finally, the procedure is applied in Section~\ref{sec6} to the numerical investigation of the elastodynamic equivalent of the Tamm problem, where fields induced by a dislocation accelerated from rest beyond the longitudinal wave speed, and thereafter put to rest again, are computed and analyzed. Section~\ref{sec7} provides a concluding discussion, which summarizes our approach and results, and points out some limitations. The most technical elements are collected in the Appendix.

\section{Basic geometric equations and field equations of motion}
\label{sec2}
\subsection{Field identities and equations of motion}
\label{sec2-1}
In this Section, the equations of motion of the elastic fields produced by moving dislocations are derived in the framework of {\it incompatible elastodynamics}~(see, e.g.,
\citet{Mura63,Mura,Kosevich,Lazar2011,Lazar2013}). An unbounded, isotropic, homogeneous, linearly elastic solid is considered. In the theory of elastodynamics of self-stresses, the equilibrium condition is\footnote{We
use the usual notation $\beta_{ij,k}:=\pd_k \beta_{ij}$ and $\dot{\beta}_{ij}:=\pd_t \beta_{ij}$.}
\begin{align}
\label{EC0}
\dot{p}_i -\sigma_{ij,j}=0\, ,
\end{align}
where $\Bp$ and $\Bsigma$ are the linear {\it momentum vector} and the
{\it stress tensor}, respectively. For incompatible linear elastodynamics, the momentum vector $\Bp$ and the stress tensor $\Bsigma$ can be expressed in terms of the
{\it elastic velocity (particle velocity) vector} $\Bv$ and the
{\it incompatible elastic distortion tensor} $\Bbeta$ by means of the two
constitutive relations
\begin{subequations}
\begin{align}
\label{CR-p}
p_i&= \rho\,  v_i\,,\\
\label{CR-t}
\sigma_{ij}&=C_{ijkl} \beta_{kl}\, ,
\end{align}
\end{subequations}
where $\rho$ denotes the mass density, and $C_{ijkl}$ the tensor of elastic moduli or elastic tensor. It enjoys the symmetry properties
$C_{ijkl}=C_{jikl}=C_{ijlk}=C_{klij}$.
For isotropic materials, the elastic tensor reduces to
\begin{align}
\label{C}
C_{ijkl}=\lambda\, \delta_{ij}\delta_{kl}
+\mu\big(\delta_{ik}\delta_{jl}+\delta_{il}\delta_{jk})\, ,
\end{align}
where $\lambda$ and $\mu$ are the Lam{\'e} constants. If the constitutive relations~(\ref{CR-p}) and (\ref{CR-t}) are substituted
into Eq.~(\ref{EC0}), the equilibrium condition expressed in terms of
the elastic fields $\Bv$ and $\Bbeta$ may be written as
\begin{align}
\label{EC}
\rho\, \dot{v}_i -C_{ijkl}\beta_{kl,j}=0\,.
\end{align}
The presence of dislocations makes the elastic fields incompatible,
which means that they are not anymore simple gradients of the material displacement vector $\Bu$.
In the eigenstrain theory of dislocations (e.g., \citet{Mura}) the total distortion tensor
$\Bbeta^\TT$ consists of elastic and
plastic parts\footnote{Note, however, that the tensors $\beta_{ij}$ and $\beta^\TP_{ij}$ defined by Mura are the transposed of the ones used in the present work. The same goes for $\alpha_{ij}$.}
\begin{align}
\label{beta-deco}
\beta^\TT_{ij}&:=u_{i,j}=\beta_{ij}+\beta^\TP_{ij},
\end{align}
but $v_i=\dot{u}_i$.
Here $\Bbeta^\TP$ is the {\it plastic distortion tensor} or
{\it eigendistortion tensor}.
The plastic distortion is a well-known quantity in dislocation theory and in Mura's theory of eigenstrain.
Nowadays, this field can be understood as a tensorial gauge field in the framework of dislocation gauge theory~\citep{LA08,Lazar10}.

For dislocations, the incompatibility tensors
are the dislocation density and
dislocation current tensors (e.g.,~\citet{Hollaender62,Kosevich,Lazar11}).
The {\it dislocation density tensor} $\Balpha$ and
the {\it dislocation current tensor} $\BI$ are classically
defined by (e.g.,~\citet{Kosevich,LL2})
\begin{subequations}
\begin{align}
\label{A2}
\alpha_{ij}&=-\epsilon_{jkl}\beta^\TP_{il,k}\,,\\
\label{I2}
I_{ij}&=-\dot{\beta}^\TP_{ij}\,,
\end{align}
\end{subequations}
or they read in terms of the elastic fields
\begin{subequations}
\begin{align}
\label{A}
\alpha_{ij}&=\epsilon_{jkl}\beta_{il,k}\,,\\
\label{I}
I_{ij}&=\dot{\beta}_{ij}-v_{i,j}\,.
\end{align}
\end{subequations}
Eqs.~(\ref{A2}) and (\ref{I2}) are the fundamental definitions of the
dislocation density tensor and of the dislocation current tensor,
respectively, whereas Eqs.~(\ref{A}) and (\ref{I}) are geometric field
identities. Originally, \citet{NYE53} introduced the concept of a dislocation
density tensor, and the definition~(\ref{A2}) of $\Balpha$ goes back to~\citet{Kroener55,Kroener58} and~\citet{Bilby55} (see also \citet{Kroener81}). The tensor $\BI$ was introduced by~\citet{Kosevich62} under the name `dislocation flux density tensor'
---a denomination used by \citet{Kosevich,Teodosiu70}, and \citet{Lardner}--- and by \citet{Hollaender62} as the `dislocation current'
(see also~\citet{Kosevich,LL2,Teodosiu70}). We adopt hereafter the latter denomination. Both $\Balpha$ and $\BI$ have nine independent components. Moreover, they fulfill the two {\it dislocation Bianchi identities} (see also~\citet{LL2,Lazar11})
\begin{subequations}
\begin{align}
\label{BI1}
\alpha_{ij,j}&=0\, ,\\
\label{BI2}
\dot{\alpha}_{ij} + \epsilon_{jkl}I_{ik,l}&= 0\, ,
\end{align}
\end{subequations}
which are geometrical consequences due to
the definitions~(\ref{A2})--(\ref{I}).
Thus,if the dislocation density tensor and dislocation current tensor are given in terms of the elastic fields and plastic fields according
to Eqs.~(\ref{A2})--(\ref{I}), then the two
dislocation Bianchi identities~(\ref{BI1}) and (\ref{BI2})
are satisfied automatically.
Conversely, if the two dislocation Bianchi identities~(\ref{BI1})
and (\ref{BI2}) are fulfilled, then the dislocation density tensor and the dislocation current tensor can be expressed in terms of elastic and plastic fields according
to Eqs.~(\ref{A2})--(\ref{I}) using the additive decomposition~(\ref{beta-deco}).
Therefore, the dislocation Bianchi identities~(\ref{BI1}) and (\ref{BI2}) are a kind of {\it compatibility conditions for the dislocation density tensor and dislocation flux tensor} or `dislocation conservation laws' (see also \citet{Kosevich}).

From the physical point of view, Eq.~(\ref{BI1}) states that dislocations do not end inside the body and Eq.~(\ref{BI2}) shows that whenever a dislocation moves or the dislocation core changes its structure and shape, the dislocation current $\BI$ is nonzero. Thus, the dislocation density can only change via the dislocation current, which means that the evolution of the dislocation density tensor $\Balpha$ is determined by the curl of the dislocation flux tensor $\BI$.

From the equilibrium condition~(\ref{EC}), uncoupled field equations for the elastic fields $\Bbeta$ and $\Bv$ produced by dislocations may be derived as equations of motion (see, e.g.,~\citet{Lazar2011,Lazar2013}). They read
\begin{subequations}
\label{Bv-NE}
\begin{align}
\label{B-NE}
L_{ik} \beta_{km}&=
\epsilon_{nml}C_{ijkl}\,\alpha_{kn,j}+\rho\,\dot{I}_{im}\, ,\\
\label{v-NE}
L_{ik} v_{k}&=C_{ijkl}\,I_{kl,j}\, ,
\end{align}
\end{subequations}
where $L_{ik}$ stands for the elastodynamic Navier differential operator
\begin{align}
\label{L}
L_{ik}=\rho\,  \delta_{ik}\pd_{tt}-C_{ijkl}\pd_j\pd_l.
\end{align}
Substituting Eq.~(\ref{C}) into Eq.~(\ref{L}), its isotropic form reads
\begin{align}
\label{GT-2D-def}
L_{ik}=\rho\,\delta_{ik} \pd_{tt}- \mu \delta_{ik}\, \Delta
-(\lambda+\mu)\, \pd_i \pd_k
\end{align}
where $\Delta$ denotes the Laplacian. Eq.~(\ref{B-NE}) is a tensorial Navier equation for $\boldsymbol{\beta}$ and Eq.~(\ref{v-NE}) is a vectorial Navier equation for $\boldsymbol{v}$, where the dislocation density and current tensors act as source terms.

\subsection{Green tensor and integral solutions}
\label{sec2-2}
We now turn to the solution of the retarded field problem of Eqs.~(\ref{B-NE}) and (\ref{v-NE}). For this purpose we use Green functions (e.g.,~\citet{Barton}). Let $\delta(.)$ denote the Dirac delta function and $\delta_{ij}$ denote the Kronecker symbol. The {\it elastodynamic Green tensor} $G^+_{ij}$ is the solution, in the sense of distributions,\footnote{The `plus' superscript serves to distinguish this distribution from the associated function $G_{ij}(\rr,t)$ to be introduced in Sec.\ \ref{sec5}.} of the (anisotropic) inhomogeneous Navier equation with unit source
\begin{align}
\label{L-iso}
L_{ik} G^+_{km}(\rr-\rr',t-t') =\delta_{im}\, \delta(t-t')\delta(\rr-\rr')\, ,
\end{align}
subjected to the causality constraint
\begin{align}
\label{G-ret}
G^+_{ij}(\rr-\rr',t-t')=0\qquad{\text{for}}\qquad t<t'\,.
\end{align}
The following properties hold in the equal-time limit (Appendix \ref{sec:etlims}):
\begin{align}
\label{G-et-lim}
\lim_{\tau\to 0^+} G^+_{ij}(\rr,\tau)=0,\qquad \lim_{\tau\to 0^+} \partial_t G^+_{ij}(\rr,\tau)=\rho^{-1}\delta_{ij}\delta(\rr).
\end{align}

Now we consider the Cauchy problem of the inhomogeneous Navier equation, expressed by Eqs.~(\ref{B-NE}) and (\ref{v-NE}). For an unbounded medium, its solutions are (see also \citet{Eringen75,Barton,Vl})
\begin{subequations}
\begin{align}
\beta_{im}(\rr,t)&=
\label{B-M0}
\epsilon_{nml}\int_{t_0}^t\d t'\int
C_{jkpl}\, G^+_{ij}(\rr-\rr', t-t')\, \alpha_{pn,k}(\rr',t')\d\rr'
\nonumber\\
&\quad
+\int_{t_0}^t\d t'\int
\rho\, G^+_{ij}(\rr-\rr', t-t')\, \dot{I}_{jm}(\rr',t')\d\rr'\nonumber\\
&\quad
+\int
G^+_{ij}(\rr-\rr', t-t_0)\, \dot{\beta}_{jm}(\rr',t_0)\,\d \rr'\nonumber\\
&\quad
+\int
\dot{G}^+_{ij}(\rr-\rr', t-t_0)\, \beta_{jm}(\rr',t_0)\, \d \rr'
\end{align}
and
\begin{align}
v_i(\rr,t)&=
\label{v-M0}
\int_{t_0}^t\d t'\int
C_{jklm}\, G^+_{ij}(\rr-\rr', t-t')\, I_{lm,k}(\rr',t')\, \d \rr'
\nonumber\\
&\quad
+\int
G^+_{ij}(\rr-\rr', t-t_0)\, \dot{v}_{j}(\rr',t_0)\, \d \rr'\nonumber\\
&\quad
+\int
\dot{G}^+_{ij}(\rr-\rr', t-t_0)\, v_{j}(\rr',t_0)\, \d \rr'\,,
\,
\end{align}
\end{subequations}
where integrals over $\rr'$ are over the whole medium, and where the following functions have been prescribed as initial conditions at $t=t_0$ throughout the medium:
\begin{align}
\label{IC}
\Bbeta(\rr,t_0)\,, \qquad \dot{\Bbeta}(\rr,t_0)\,,\qquad
\Bv(\rr,t_0)\,,\qquad \dot{\Bv}(\rr,t_0)\,.
\end{align}
Because the elastodynamic Navier equation is a generalization of the wave equation, Eqs.~(\ref{B-M0}) and (\ref{v-M0}) are similar to the Poisson formula for the latter~\citep{Vl}.

Since $G^+_{ij}(\rr-\rr', t-t_0)$ and $\dot{G}^+_{ij}(\rr-\rr', t-t_0)$ vanish as $t_0\to-\infty$, Eqs.~(\ref{B-M0}) and (\ref{v-M0}) can be represented as convolutions of the Green tensor with the sources of the inhomogeneous Navier equations, only~\citep{Mura63,Lazar2011}. Letting thus $t_0\rightarrow-\infty$ the solutions for $\Bbeta$ and $\Bv$ reduce to
\begin{subequations}
\begin{align}
\beta_{im}(\rr,t)&=
\label{B-M}
\epsilon_{nml}\int_{-\infty}^t\d t'\int
C_{jkpl}\, G^+_{ij}(\rr-\rr', t-t')\, \alpha_{pn,k}(\rr',t')\, \d \rr'
\nonumber\\
&\quad
+\int_{-\infty}^t \d t'\int
\rho\, G^+_{ij}(\rr-\rr', t-t')\, \dot{I}_{jm}(\rr',t')\, \d \rr'\,,\\
v_i(\rr,t)&=
\label{v-M}
\int_{-\infty}^t\d t'\int
C_{jklm}\, G^+_{ij}(\rr-\rr', t-t')\, I_{lm,k}(\rr',t')\, \d \rr'\,,
\end{align}
\end{subequations}
or equivalently
\begin{subequations}
\begin{align}
\beta_{im}(\rr,t)&=
\label{B-M-1}
\epsilon_{nml}\int_{-\infty}^t\d t'\int
C_{jkpl}\, G^+_{ij,k}(\rr-\rr', t-t')\, \alpha_{pn}(\rr',t')\, \d \rr'
\nonumber\\
&\quad
+\int_{-\infty}^t\d t'\int
\rho\, \dot{G}^+_{ij}(\rr-\rr', t-t')\, I_{jm}(\rr',t')\, \d \rr'\,,\\
v_i(\rr,t)&=
\label{v-M-1}
\int_{-\infty}^t\d t'\int
C_{jklm}\, G^+_{ij,k}(\rr-\rr', t-t')\, I_{lm}(\rr',t')\, \d \rr'\,.
\end{align}
\end{subequations}
Eqs.~(\ref{B-M})--(\ref{v-M-1}) are valid for general dislocation distributions (continuous distribution of dislocations, dislocation loops, straight dislocations). Later on, we shall specialize to straight dislocations.

Let the velocity $\BV(t)$ of a moving dislocation be some given function of time. Then, the following relation holds between its associated dislocation density and current tensors:
\begin{align}
\label{rel-V}
I_{ij}=\epsilon_{jkn}\, V_k\, \alpha_{in}\,.
\end{align}
This relation means that the current $\BI$ is caused by the moving dislocation density $\Balpha$. Thus, $\BI$ is a convection dislocation current \citep{Guenther73,Lazar2013}. Substituting Eq.~(\ref{rel-V}) into relation~(\ref{BI2}), the Bianchi identity~(\ref{BI2}) reduces to the following form in terms of the dislocation density tensor and the dislocation velocity vector
\begin{align}
\label{BI3}
\dot{\alpha}_{ij} = -\epsilon_{jkl}(\epsilon_{kmn} V_m \alpha_{in})_{,l}
=(V_j \alpha_{il})_{,l}-(V_l \alpha_{ij})_{,l}\,.
\end{align}
Sometimes the Bianchi identity~(\ref{BI3}) is called {\it dislocation density transport equation} (see, e.g.,~\citet{Djaka15}).

We moreover obtain from Eqs.~(\ref{B-NE}) and (\ref{v-NE}) the field equations of motion in the form
\begin{subequations}
\begin{align}
\label{B-NE-2}
L_{ik}
\beta_{km}&=
\epsilon_{nml}
\big[C_{ijkl}\,\alpha_{kn,j}+\rho\,\pd_t(V_l\alpha_{in})\big]
\, ,\\
\label{v-NE-2}
L_{ik}
v_{k}&=\epsilon_{nml}C_{ijkm}\,(V_l \alpha_{kn})_{,j}\, ,
\end{align}
\end{subequations}
where the sources are given in terms of the dislocation density tensor and the dislocation velocity vector. Obviously, the validity of Eq.\ (\ref{BI3}) is conditioned by the assumptions that underlie Eq.\ (\ref{rel-V}). Thus, Eq.~(\ref{rel-V}) makes sense only for a discrete dislocation line with rigid core, since it neglects changes with time of its core shape. However, by imposing a suitable parameterization of the dislocation density or of the plastic eigenstrain (e.g., \cite{PELL14}), an additional term in the current tensor related to core-width variations could easily be derived from Eq.~(\ref{I2}). \emph{Such effects are neglected in the present study.}
Accordingly, from Eq.~(\ref{BI3}) and using $V_{j,l}=V_{l,l}=0$ for the problem considered, we deduce with the help of the Bianchi identity~(\ref{BI1}) that for one single rigid dislocation
\begin{align}
\label{eq:alphadotsingle}
\dot{\alpha}_{ij}&=- V_k\,\alpha_{ij,k}.
\end{align}

\subsection{Remarks}
\label{sec:remarks}
It is worthwhile pointing out that although we insisted, for better physical insight, on deriving Eq.\ (\ref{B-M-1}) from field equations with sources expressed in terms of dislocation density and current, the latter equation is fully consistent with the perhaps more familiar writing of the elastic distortion in terms of the plastic distortion and the second derivatives of the Green tensor as (e.g., \cite{Mura})\footnote{
Indeed, ignoring our present emphasis on the distributional character of the Green tensor, Eq.\ (\ref{B-M-1}) is nothing but Eq.\ (38.36) on p.\ 351 of Mura's treatise. This is realized upon comparing Mura's Eq.\ (38.19) with the above definition (\ref{I2}) of $I_{ij}$, bearing in mind the transposed character of our dislocation tensors with respect to Mura's (see note 3).}
\begin{align}
\label{eq:distfam}
\beta_{ij}(\rr,t)&=-\int_{-\infty}^t\d t'\int\d\rr' G^+_{ik,jl}(\rr-\rr',t-t')C_{klmn}\beta^\TP_{mn}(\rr',t')-\beta^\TP_{ij}(\rr,t)\,.
\end{align}

Also, the issue of the upper boundary $t'=t$ of the time-integration in the integral solutions deserves some comments. It is sometimes read in treatises on Green functions, e.g., \citep{Barton}, that the upper boundary should lie slightly above $t$, which is usually denoted by $t^+$. Such a device helps one to easily check that the integral formulas are indeed solutions of the equation of motion they derive from. Because of the causality constraint, the upper time-integration boundary can as well be taken as $+\infty$. However, it is less recognized that the boundary can as well be chosen slightly \emph{below} $t$, which we denote as $t^-$. This is possible because of the two limiting properties (\ref{G-et-lim}), the first of which ensuring that removing the interval $]t^-,t^+[$ from the integration interval $t'\in ]-\infty,t^+[$ makes no difference on the final result. The second property in (\ref{G-et-lim}) allows us to show ---in Appendix \ref{sec:tminus}--- that solutions written with integrals over $t'\in ]-\infty,t^-[$ satisfy the equation of motion as well. Since the solution is unique, all these formulations give identical results. However, the use of $t^-$, which amounts to eliminating the immediate vicinity of the point $t'=t$ from time integrals, is much more convenient for numerical and analytical purposes, as will be shown in Section \ref{sec:sec5-2}. This device has already been employed in \citep{Pellegrini11,Pellegrini12,PELL14,PL2014}, but was introduced there without any detailed justification. Until Section \ref{sec4}, we continue denoting the upper boundary as $t'=t$ in general formulas, for simplicity.

\section{Straight Volterra dislocations in the framework of distributions}
\label{sec3}
In this Section, the elastodynamic fields produced by the non-uniform motion of straight screw and edge Volterra dislocations are studied using the theory of distributions or generalized functions~\citep{Schwartz,GS,Kanwal}.
The field equations of motion are solved by means of Green functions. The problem is two-dimensional, of anti-plane strain or plane strain character.

\subsection{Screw dislocation}
\label{sec3-1}
We address first the anti-plane strain problem of a Volterra screw dislocation in non-uniform motion at time $t$ along some arbitrary path $\Bs(t')$ prescribed in advance in the time range $-\infty<t'\leq t$. The dislocation line and the Burgers vector $b_z$ are parallel to the $z$-axis. The dislocation velocity has two non-vanishing components: $V_x=\dot{s}_x(t)$, $V_y=\dot{s}_y(t)$. The dislocation density and dislocation current tensors are
\begin{align}
\label{dd-s}
\alpha_{zz}=b_z\, \ell_z\, \delta(\BR(t))\, ,\qquad
I_{zj}=b_z\, \epsilon_{jkz} V_k(t) \,\ell_z\, \delta(\BR(t))\, ,
\end{align}
where $\BR(t)=\rr-\Bs(t) \in \R^2$,
$\ell_z$ is a unit vector in $z$-direction
 and $i,j,k=x,y$. The index $z$ is a fixed index (no summation).

Eqs.~(\ref{B-NE-2}) and (\ref{v-NE-2}) simplify enormously for the nonvanishing components $\beta_{zx}$, $\beta_{zy}$, and $v_z$. Using Eq.~(\ref{dd-s}) and $\dot{\alpha}_{zz}=-V_k \alpha_{zz,k}$, we obtain from Eqs.~(\ref{B-NE-2}) and (\ref{v-NE-2}) the following equations of motion of a screw dislocation:
\begin{subequations}
\begin{align}
\label{B-NE-s}
L_{zz}\beta_{zm}&=\epsilon_{zml}\big[C_{zjzl}\,\alpha_{zz,j}+\rho\,\big(\dot{V}_l\,\alpha_{zz}-V_l V_k\,\alpha_{zz,k}\big)\big]\, ,\\
\label{v-NE-s}
L_{zz}
v_{z}&=\epsilon_{zml}\,C_{zjzm}\,V_l\,\alpha_{zz,j}\, ,
\end{align}
\end{subequations}
where $\dot{\BV}$ is the dislocation acceleration. With the dynamic elastic tensor for non-uniform motion, namely,
\begin{align}
\label{C-dyn}
\widetilde{C}_{ijkl}(\BV)&=C_{ijkl}-\rho\,V_j V_l\, \delta_{ik}
\end{align}
and using the property of the differentiation of a convolution, the appropriate solution may be written as the convolution integrals
\begin{subequations}
\begin{align}
\beta_{zm}(\rr,t)&=
\label{B-s-1}
\epsilon_{zml}\int_{-\infty}^t\d t'\int\Big\{G^+_{zz,k}(\rr-\rr', t-t')\,\widetilde{C}_{zkzl}(\BV(t'))\nonumber\\
&\qquad\qquad\qquad\qquad
+ \rho\, G^+_{zz}(\rr-\rr', t-t')\, \dot{V}_l(t')\Big\}\,\alpha_{zz}(\rr',t')\,\d \rr'\,,
\\
v_z(\rr,t)&=
\label{v-s-1}
\epsilon_{zml}\int_{-\infty}^t\d t'\int
C_{zkzm}\, G^+_{zz,k}(\rr-\rr', t-t')\, V_l(t')\, \alpha_{zz}(\rr',t')
\, \d \rr'\, ,
\end{align}
\end{subequations}
The dynamic elastic tensor (\ref{C-dyn}) was originally introduced by \citet{Saenz} for uniformly moving dislocations (see also \citet{Bacon}, who use a different index ordering), and employed in elastodynamics by \citet{Wu} with the same index ordering as in Eq.\ (\ref{C-dyn}). It possesses only the major symmetry $\widetilde{C}_{ijkl}(\BV)= \widetilde{C}_{klij}(\BV)$.

Substituting the dislocation density~(\ref{dd-s}) into Eqs.~(\ref{B-s-1}) and (\ref{v-s-1}), and performing the integration over $\rr'$, we obtain
\begin{subequations}
\label{eq:voltintscrew}
\begin{align}
\beta_{zm}(\rr,t)&=
\label{B-z}
 b_{z}\ell_z\,
\epsilon_{zml}\int_{-\infty}^t
\Big\{
G^+_{zz,k}(\rr-\Bs(t'), t-t')\,\widetilde{C}_{zkzl}(\BV(t'))+\rho\, G^+_{zz}(\rr-\Bs(t'), t-t')\,\dot{V}_l(t')
\Big\} \d t'\\
\label{v-3}
v_z(\rr,t)&=b_{z}\ell_z\,\epsilon_{zml}\int_{-\infty}^t C_{zkzm}\, G^+_{zz,k}(\rr-\Bs(t'), t-t')\, V_l(t')\, \d t'\, ,
\end{align}
\end{subequations}
where $G^+_{zz}$ is the retarded Green function (distribution) of the anti-plane problem defined by
\begin{align}
\label{GT-2D-zz}
L_{zz}G^+_{zz}=
\big(\rho \, \pd_{tt}- \mu\, \Delta\big)G^+_{zz}=\delta(t)\delta(\rr)\, .
\end{align}

If the material is infinitely extended, the two-dimensional elastodynamic Green-function distribution of the anti-plane problem, which is nothing but the usual Green function of the two-dimensional scalar wave equation (e.g.~\citet{MF,Barton}), interpreted as a distribution, reads~(see, e.g., \citet{Eringen75,Kausel})
\begin{align}
\label{GT-zz}
G^+_{zz}(\rr,t)=
\frac{\theta(t)}{2\pi\mu}\,
\big(t^2-r^2/c^2_\TT\big)^{-1/2}_{+}\,
\end{align}
with the velocity of transverse elastic waves (shear waves, also called S-waves)
\begin{align}
\label{cT}
c_{\TT}=\sqrt{\mu/\rho}\,,
\end{align}
and where $\theta(t)$ is the Heaviside unit-step function that restricts this causal solution to positive times. In this writing, the generalized function $x^\lambda_+$, defined as
(see, e.g., \citet{Schwartz,GS,Kanwal,deJager})
\begin{align}
\label{x+}
x^\lambda_+=
\left\{
\begin{array}{ll}
\displaystyle{0}\quad &
\displaystyle{\text{for}\ x<0}
\\
\displaystyle{x^\lambda}\quad &
\displaystyle{\text{for}\  x>0}\ , \\
\end{array}
\right.
\end{align}
has been used.
The derivative of $x^{1/2}_+$ is given by
\begin{align}
\label{x+_g1}
\big(x^{1/2}_+\big)'=\frac{1}{2}\, x^{-1/2}_+\,.
\end{align}
The derivative of $x^{-1/2}_+$ gives a pseudofunction
(see~\citet{Schwartz,GS,Zemanian}):
\begin{align}
\label{x+_g2}
\big(x^{-1/2}_+\big)'=-\frac{1}{2}\, \text{Pf}\ x^{-3/2}_+ \, .
\end{align}
The symbol Pf in Eq.~(\ref{x+_g2}) stands for pseudofunction. In general, pseudofunctions are distributions generated by Hadamard's finite part of a divergent integral. They arise naturally when certain distributions are differentiated. In Eq.~(\ref{x+_g2}) the regular distribution $x^{-1/2}_+$ was differentiated. Using Eqs.~(\ref{x+_g1}) and (\ref{x+_g2}), the derivative of the Green function~(\ref{GT-zz}) is expressed as the pseudofunction
\begin{align}
\label{GT-zz-grad}
G^+_{zz,k}(\rr,t)=
\frac{\theta(t)}{2\pi\mu}\,
\frac{x_k}{c_\TT^2}\,
\text{Pf}\, \big(t^2-r^2/c^2_\TT\big)^{-3/2}_{+}\,.
\end{align}

Finally, the elastic fields of a non-uniformly moving screw Volterra dislocation read, in distributional form
\begin{subequations}
\label{eq:bziv}
\begin{align}
\beta_{zx}(\rr,t)&=
\label{Bzx-s}
\frac{b_z\ell_z}{2\pi c_\TT^2}
\int_{-\infty}^t
\bigg(
\dot{V}_y(t')\, \big[\bar{t}^2-R^2(t')/c_\TT^2\big]^{-1/2}_{+}
\nonumber\\
&\qquad
+\bigg(\bigg(1-\frac{V^2_y(t')}{c^2_{\TT}}\bigg) R_y(t')
-\frac{V_x(t')V_y(t')}{c^2_{\TT}}\, R_x(t')\bigg)\,
\text{Pf}\, \big[\bar{t}^2-R^2(t')/c_\TT^2\big]^{-3/2}_{+}
\bigg)\d t'\,,\\
\beta_{zy}(\rr,t)&=
\label{Bzy-s}
-\frac{b_z\ell_z}{2\pi c_\TT^2}
\int_{-\infty}^t
\bigg(
\dot{V}_x(t')\, \big[\bar{t}^2-R^2(t')/c_\TT^2\big]^{-1/2}_{+}
\nonumber\\
&\qquad
+\bigg(\bigg(1-\frac{V^2_x(t')}{c^2_{\TT}}\bigg) R_x(t')
-\frac{V_x(t')V_y(t')}{c^2_{\TT}}\, R_y(t')\bigg)\,
\text{Pf}\, \big[\bar{t}^2-R^2(t')/c_\TT^2\big]^{-3/2}_{+}
\bigg)\d t'\,,\\
v_z(\rr,t)&=
\label{v-s}
\frac{b_z\ell_z}{2\pi c_\TT^2}\int_{-\infty}^t
\big(V_y(t') R_x(t')-V_x(t')R_y(t')\big)\,
\text{Pf}\, \big[\bar{t}^2-R^2(t')/c_\TT^2\big]^{-3/2}_{+}
\, \d t'\, ,
\end{align}
\end{subequations}
where $\bar{t}=t-t'$.

The fields given by Eqs.~(\ref{Bzx-s})--(\ref{v-s}) clearly consist of two parts:
(i) Fields depending on the dislocation velocities $V_x$ and $V_y$ alone and proportional to the pseudofunction distribution of power $-3/2$ ---dislocation velocity-dependent fields or \emph{near fields}, built from the gradient of the Green tensor;
(ii) Fields depending on the dislocation accelerations $\dot{V}_x$ and $\dot{V}_y$ and proportional to the regular distribution of power $-1/2$ ---dislocation acceleration-dependent fields or \emph{far fields},
built on the Green tensor itself. The velocity field~(\ref{v-s}) possesses no acceleration part.

It should be mentioned that it seems to be hard to find a measurement which can distinguish between the acceleration- and velocity-depending fields. Such a decomposition is basically conceptual. In a natural way, we may separate $\Bbeta$ into two parts, one which involves the dislocation acceleration and goes to zero for $\dot{\BV}=0$, and one which involves only the dislocation velocity and yields the static field for a dislocation with $\BV=0$. Dislocations at rest or in steady motion do not generate elastodynamic waves. Only non-uniformly moving dislocations emit elastodynamic radiation.

Some historical remarks are in order. In the 1950s already, \citet{Sauer1,Sauer2} emphasized the interest of introducing the theory of distributions in supersonic aerodynamics. In particular, in gas dynamics and wing theory, pseudofunctions of power $-3/2$ have been used in the framework of distribution theory, e.g., \citet{Sauer1,Sauer2,Dorfner} (see also~\citet{deJager}).

\subsection{Edge dislocation}
\label{sec3-2}
We next turn to the straight edge Volterra dislocation in the plane-strain framework. Its associated dislocation density and current tensors read, respectively,
\begin{align}
\label{dd-e}
\alpha_{ij}=b_i\, \ell_j\, \delta(\BR(t))\, ,\qquad
I_{ij}=b_i\, \epsilon_{jkl} V_k(t)\, \ell_l\, \delta(\BR(t))\, ,
\end{align}
where $\BR(t)=\rr-\Bs(t) \in \R^2$ and $i,j,k=x,y$.
Using $\dot{\alpha}_{ij}=-V_k\,\alpha_{ij,k}$,
we obtain from Eqs.~(\ref{B-NE-2}) and (\ref{v-NE-2}) the following equations of motion:
\begin{subequations}
\begin{align}
\label{B-NE-3}
L_{ik} \beta_{km}&=
\epsilon_{nml}\big[
C_{ijkl}\,\alpha_{kn,j}+\rho\,\big( \dot{V}_l\alpha_{in}-V_l V_k\,\alpha_{in,k}\big)\big]\, ,\\
\label{v-NE-3}
L_{ik}v_{k}&=\epsilon_{nml}\, C_{ijkl}\,V_l\,\alpha_{kn,j}\, ,
\end{align}
\end{subequations}
where $\BV=\dot{\Bs}$. Using the property of the differentiation of a convolution, the corresponding solutions of Eqs.~(\ref{B-NE-3}) and (\ref{v-NE-3}) are given in convolution form
\begin{subequations}
\begin{align}
\beta_{im}(\rr,t)&=
\label{B-1}
\epsilon_{nml}\int_{-\infty}^t\d t'\int
\Big(G^+_{ij,k}(\rr-\rr', t-t')\,\widetilde{C}_{jkpl}(\BV(t'))\, \alpha_{pn}(\rr',t')
\nonumber\\
&\qquad\qquad\qquad\qquad
+ \rho\, G^+_{ij}(\rr-\rr', t-t')\, \dot{V}_l(t') \alpha_{jn}(\rr',t')\Big)\d \rr'\,,\\
\label{v-1}
v_i(\rr,t)&=\epsilon_{nml}\int_{-\infty}^t\d t'\int C_{jkpm}\, G^+_{ij,k}(\rr-\rr', t-t')\, V_l(t') \alpha_{pn}(\rr',t')\, \d \rr'\,,
\end{align}
\end{subequations}
where the two-dimensional (distributional) Green tensor $G^+_{ij}$ is defined as the retarded solution of Eq.~(\ref{L-iso}), with (\ref{GT-2D-def}).

Substituting Eq.~(\ref{dd-e}) into (\ref{B-1}) and (\ref{v-1}) and
performing the $\rr'$-integration, we find
\begin{subequations}
\label{eq:voltintedge}
\begin{align}
\beta_{im}(\rr,t)&=
\label{B-e}
\epsilon_{nml}\int_{-\infty}^t\Big(G^+_{ij,k}(\rr-\Bs(t'), t-t')\,\widetilde{C}_{jkpl}(\BV(t'))\, b_p \ell_n
\nonumber\\
&\qquad\qquad\qquad\qquad
+ \rho\, G^+_{ij}(\rr-\Bs(t'), t-t')\, \dot{V}_l(t')\, b_{j}\ell_n\Big) \d t'\\
v_i(\rr,t)&=
\label{v-e}
\epsilon_{nml}\int_{-\infty}^t C_{jkpm}\, G^+_{ij,k}(\rr-\Bs(t'), t-t')\, V_l(t')\, b_{p}\ell_n\, \d t'\, .
\end{align}
\end{subequations}

Using the distributional approach, the two-dimensional retarded Green tensor is given by (see \citet{EASO56,Eringen75,Kausel} for the Green tensor in the classical approach)
\begin{align}
G^+_{ij}(\rr,t)&=\frac{\theta(t)}{2\pi\rho }\,
\bigg\{
\frac{x_i x_j}{r^4}\,
\Big[
t^2\, \big(t^2-r^2/c^2_\TL\big)^{-1/2}_{+}
+\big(t^2-r^2/c^2_\TL\big)^{1/2}_{+}
-t^2\, \big(t^2-r^2/c^2_\TT\big)^{-1/2}_{+}
-\big(t^2-r^2/c^2_\TT\big)^{1/2}_{+}
\Big]
\nonumber\\
\label{GT-2D}
&\qquad\qquad
-\frac{\delta_{ij}}{r^2}\,
\Big[
\big(t^2-r^2/c^2_\TL\big)^{1/2}_{+}
-t^2\, \big(t^2-r^2/c^2_\TT\big)^{-1/2}_{+}
\Big]\bigg\}\, .
\end{align}
It consists of regular distributions of power $1/2$ and $-1/2$. The shear velocity $c_\TL$ is defined in (\ref{cT}), and $c_\TT$ is the velocity of the longitudinal elastic waves (P-wave) expressed in terms of the Lam\'e constants as
\begin{align}
\label{c}
c_{\TL}=\sqrt{(2\mu+\lambda)/\rho}.
\end{align}
It is noted that $G_{zz}^+(\rr,t)$, Eq.\ (\ref{GT-zz}), is twice the spherical part of $G_{ij}^+(\rr,t)$ in (\ref{GT-2D}). Using Eqs.~(\ref{x+_g1}) and (\ref{x+_g2}), the derivative of the Green tensor~(\ref{GT-2D}) is obtained as
\begin{align}
\label{GT-2D-grad}
G^+_{ij,k}(\rr,t)&=\frac{\theta(t)}{2\pi\rho }\,
\bigg\{
\bigg(
\frac{\delta_{ik} x_j+\delta_{jk} x_i }{r^4}-\frac{4\, x_i x_jx_k}{r^6}
\bigg)
\Big[
t^2\, \big(t^2-r^2/c^2_\TL\big)^{-1/2}_{+}
+\big(t^2-r^2/c^2_\TL\big)^{1/2}_{+}
\nonumber\\
&\hspace{6.3cm}
-t^2\, \big(t^2-r^2/c^2_\TT\big)^{-1/2}_{+}
-\big(t^2-r^2/c^2_\TT\big)^{1/2}_{+}
\Big]
\nonumber\\
&\qquad\qquad
+\frac{2\,\delta_{ij} x_k}{r^4}\,
\Big[
\big(t^2-r^2/c^2_\TL\big)^{1/2}_{+}
-t^2\, \big(t^2-r^2/c^2_\TT\big)^{-1/2}_{+}
\Big]
\nonumber\\
&\qquad\qquad
+\frac{x_i x_j x_k }{r^4}\,
\bigg[
\frac{t^2}{c_\TL^2}\, \text{Pf}\,\big(t^2-r^2/c^2_\TL\big)^{-3/2}_{+}
-\frac{1}{c_\TL^2}\big(t^2-r^2/c^2_\TL\big)^{-1/2}_{+}
\nonumber\\
&\qquad\qquad\qquad\qquad
-\frac{t^2}{c_\TT^2}\, \text{Pf}\,\big(t^2-r^2/c^2_\TT\big)^{-3/2}_{+}
+\frac{1}{c_\TT^2}\big(t^2-r^2/c^2_\TT\big)^{-1/2}_{+}
\bigg]
\nonumber\\
&\qquad\qquad
+\frac{\delta_{ij}x_k }{r^2}\,
\bigg[\frac{1}{c_\TL^2}\big(t^2-r^2/c^2_\TL\big)^{-1/2}_{+}
+\frac{t^2}{c_\TT^2}\, \text{Pf}\,\big(t^2-r^2/c^2_\TT\big)^{-3/2}_{+}
\bigg]
\bigg\}\,,
\end{align}
which involves pseudofunctions of power $-3/2$ in addition to distributions of power $1/2$ and $-1/2$. Eqs.\ (\ref{GT-2D}) and (\ref{GT-2D-grad}) display $G^+_{ij}(\rr,t)$ and $G^+_{ij,k}(\rr,t)$ in expanded form for clarity. However, more compact expressions for these distributions that emphasize the occurrence of $t$ solely via well-defined groups containing either $\cT t$ or $\cL t$ can be found in \citep{PL2014} [see also Eq.\ (\ref{eq:greencompact}) below].

Substituting Eqs.~(\ref{GT-2D}) and (\ref{GT-2D-grad}) into Eqs.~(\ref{B-e}) and (\ref{v-e}), we obtain the elastic distortion tensor of the non-uniformly moving Volterra edge dislocation as
\begin{align}
\beta_{im}(\rr,t)&=
\label{B-2}
\frac{1}{2\pi\rho }\,\epsilon_{nml}\int_{-\infty}^t  \d t'
\bigg(\widetilde{C}_{jkpl}(\BV(t'))\, b_p \ell_n\, \bigg\{
\bigg(
\frac{\delta_{ik} R_j(t')+\delta_{jk} R_i(t') }{R^4(t')}
-\frac{4\, R_i(t') R_j(t') R_k(t')}{R^6(t')}
\bigg)\nonumber\\
&
\qquad\qquad\qquad\qquad
\times
\bigg[
\tt^2\, \Big(\tt^2-R^2(t')/c^2_\TL\Big)^{-1/2}_{+}
+\Big(\tt^2-R^2(t')/c^2_\TL\Big)^{1/2}_{+}
\nonumber\\
&
\qquad\qquad\qquad\qquad
-\tt^2\, \Big(\tt^2-R^2(t')/c^2_\TT\Big)^{-1/2}_{+}
-\Big(\tt^2-R^2(t')/c^2_\TT\Big)^{1/2}_{+}
\Bigg]
\nonumber\\
&\qquad\qquad
+\frac{2\,\delta_{ij} R_k(t')}{R^4(t')}\,
\Big[
\big(\tt^2-R^2(t')/c^2_\TL\big)^{1/2}_{+}
-\tt^2\, \big(\tt^2-R^2(t')/c^2_\TT\big)^{-1/2}_{+}
\Big]
\nonumber\\
&\qquad\qquad
+\frac{R_i(t') R_j(t') R_k(t') }{R^4(t')}\,
\bigg[
\frac{\tt^2}{c_\TL^2}\, \text{Pf}\,\big(\tt^2-R^2(t')/c^2_\TL\big)^{-3/2}_{+}
-\frac{1}{c_\TL^2}\big(\tt^2-R^2(t')/c^2_\TL\big)^{-1/2}_{+}
\nonumber\\
&\qquad\qquad\qquad\qquad
-\frac{\tt^2}{c_\TT^2}\, \text{Pf}\,\big(\tt^2-R^2(t')/c^2_\TT\big)^{-3/2}_{+}
+\frac{1}{c_\TT^2}\big(\tt^2-R^2(t')/c^2_\TT\big)^{-1/2}_{+}
\bigg]
\nonumber\\
&\qquad\qquad
+\frac{\delta_{ij}R_k(t') }{R^2(t')}\,
\bigg[\frac{1}{c_\TL^2}\big(\tt^2-R^2(t')/c^2_\TL\big)^{-1/2}_{+}
+\frac{\tt^2}{c_\TT^2}\, \text{Pf}\,\big(\tt^2-R^2(t')/c^2_\TT\big)^{-3/2}_{+}
\bigg]
\bigg\}\,
\nonumber\\
&\qquad\qquad
+ \rho\,\dot{V}_l(t')\,  b_{j} \ell_n
\bigg\{\frac{R_i(t') R_j(t')}{R^4(t')}\,
\Big[
\tt^2\, \big(\tt^2-R^2(t')/c^2_\TL\big)^{-1/2}_{+}
+\big(\tt^2-R^2(t')/c^2_\TL\big)^{1/2}_{+}
\nonumber\\
&\qquad\qquad\qquad\qquad\qquad
-\tt^2\, \big(\tt^2-R^2(t')/c^2_\TT\big)^{-1/2}_{+}
-\big(\tt^2-R^2(t')/c^2_\TT\big)^{1/2}_{+}
\Big]
\nonumber\\
&\qquad\qquad
-\frac{\delta_{ij}}{R^2(t')}\,
\Big[
\big(\tt^2-R^2(t')/c^2_\TL\big)^{1/2}_{+}
-\tt^2\, \big(\tt^2-R^2(t')/c^2_\TT\big)^{-1/2}_{+}
\Big]\bigg\}\bigg)\,,
\end{align}
and the velocity vector reads
\begin{align}
v_i(\rr,t)&=
\label{v-2}
\frac{ b_{p}\ell_n}{2\pi\rho }\, \epsilon_{nml}\,C_{jkpm}
\int_{-\infty}^t \d t'\,
V_l(t')\,
\bigg\{
\bigg(
\frac{\delta_{ik} R_j(t')+\delta_{jk} R_i(t') }{R^4(t')}
-\frac{4\, R_i(t') R_j(t') R_k(t')}{R^6(t')}
\bigg)\nonumber\\
&
\qquad\qquad\qquad\qquad
\times
\bigg[
\tt^2\, \Big(\tt^2-R^2(t')/c^2_\TL\Big)^{-1/2}_{+}
+\Big(\tt^2-R^2(t')/c^2_\TL\Big)^{1/2}_{+}
\nonumber\\
&
\qquad\qquad\qquad\qquad
-\tt^2\, \Big(\tt^2-R^2(t')/c^2_\TT\Big)^{-1/2}_{+}
-\Big(\tt^2-R^2(t')/c^2_\TT\Big)^{1/2}_{+}
\Bigg]
\nonumber\\
&\qquad\qquad
+\frac{2\,\delta_{ij} R_k(t')}{R^4(t')}\,
\Big[
\big(\tt^2-R^2(t')/c^2_\TL\big)^{1/2}_{+}
-\tt^2\, \big(\tt^2-R^2(t')/c^2_\TT\big)^{-1/2}_{+}
\Big]
\nonumber\\
&\qquad\qquad
+\frac{R_i(t') R_j(t') R_k(t') }{R^4(t')}\,
\bigg[
\frac{\tt^2}{c_\TL^2}\, \text{Pf}\,\big(\tt^2-R^2(t')/c^2_\TL\big)^{-3/2}_{+}
-\frac{1}{c_\TL^2}\big(\tt^2-R^2(t')/c^2_\TL\big)^{-1/2}_{+}
\nonumber\\
&\qquad\qquad\qquad\qquad
-\frac{\tt^2}{c_\TT^2}\, \text{Pf}\,\big(\tt^2-R^2(t')/c^2_\TT\big)^{-3/2}_{+}
+\frac{1}{c_\TT^2}\big(\tt^2-R^2(t')/c^2_\TT\big)^{-1/2}_{+}
\bigg]
\nonumber\\
&\qquad\qquad
+\frac{\delta_{ij}R_k(t') }{R^2(t')}\,
\bigg[\frac{1}{c_\TL^2}\big(\tt^2-R^2(t')/c^2_\TL\big)^{-1/2}_{+}
+\frac{\tt^2}{c_\TT^2}\, \text{Pf}\,\big(\tt^2-R^2(t')/c^2_\TT\big)^{-3/2}_{+}
\bigg]
\bigg\}\,.
\end{align}
Those elastic fields consist of two different kinds of contributions, about which the same comments as in the screw case can be made.

Expressions~(\ref{B-2})--(\ref{v-2}) encompass gliding as well as climbing edge dislocations. If $\BV\| \Bb$, they describe a gliding edge dislocation, and with $\BV\perp \Bb$  they deliver the fields of a climbing edge dislocation (see, e.g., \citet{Lazar2011,Pellegrini10}). If we specialize to non-uniformly moving straight edge dislocations with Burgers vector in the $x$-direction, $b_x$, and with the dislocation line $\ell_z$ parallel to the $z$-axis, then the dislocation density and current tensors of a gliding edge dislocation with arbitrary velocity $V_x(t)$ in the $x$-direction are given by
\begin{align}
\label{dd-gl}
\alpha_{xz}=b_x\, \ell_z\, \delta(\BR(t))\, ,\qquad
I_{xy}=b_x\, \epsilon_{yxz} V_x(t)\, \ell_z\, \delta(\BR(t))\, ,
\end{align}
where $\BR(t)=(x-s_x(t),y)$.
For a climbing edge dislocation with arbitrary velocity $V_y(t)$ in the $y$-direction, they read
\begin{align}
\label{dd-cl}
\alpha_{xz}=b_x\, \ell_z\, \delta(\BR(t))\, ,\qquad
I_{xx}=b_x\, \epsilon_{xyz} V_y(t)\, \ell_z\, \delta(\BR(t))\, ,
\end{align}
where $\BR(t)=(x,y-s_y(t))$.

In general, the elastodynamic fields of straight dislocations have the form of time-integrals over the history of the motion, and display a so-called `afterglow'-type response  \citep{Barton} with slow relaxation tails. The reason, specific to the two-dimensional problem, is that fields continuously arrive from remote emission points on past locations of the dislocation line. This effect is accounted for by the two-dimensional Green function.

We are therefore left to evaluate time-integrals of considerable complexity, which only in some simple cases yield closed-form results in terms of elementary functions.
Consequently, the procedure developed hereafter relies, after a suitable regularization method has been applied, on a decomposition of arbitrary motion into time-intervals of constant velocity for which explicit field expressions can be given.

\section{Regularization in the framework of distributions}
\label{sec4}
\subsection{Regularization procedure}
\label{sec4-1}
Up to now, our results for the fields of non-uniformly moving Volterra dislocations are singular distributions. Although being mathematically well-defined in the latter sense, and therefore free of non-integrable singularities, they are inconvenient for numerical purposes in the case of arbitrarily prescribed motion $\Bs(t)$. In order to get singularity-free fields, we have to regularize these distributions. The standard means of doing this is the convolution of distributions with a suitable test function. This operation, which is called the {\it regularization of a distribution}, converts the distribution into an infinitely smooth function.

\begin{figure}
\begin{centering}
\includegraphics[width=8.5cm]{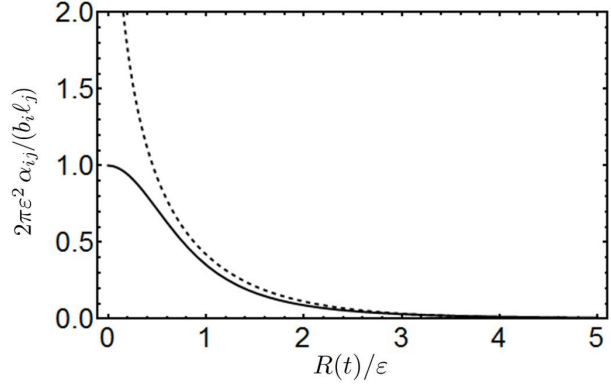}
\caption{\label{fig:alpha}
Scaled dislocation densities versus scaled distance to core center: regularized dislocation density $\alpha^{\text{iso}}_{ij}$ (solid), and dislocation density $\alpha_{ij}$ from gradient elasticity of Helmholtz type (dashed).}
\end{centering}
\end{figure}

The procedure we call hereafter \emph{isotropic regularization} consists in the convolution of the singular distributions by the following isotropic representation of the two-dimensional Dirac delta distribution \citep{Kanwal}:
\begin{align}
\label{reg-f}
\delta(\rr)=\delta(x)\delta(y)=\lim_{\varepsilon\to 0}
\delta^\varepsilon(\rr)\,,\qquad\text{with}\qquad
\delta^\varepsilon(\rr)=\frac{\varepsilon}{2\pi(r^2+\varepsilon^2)^{3/2}}\,,
\end{align}
which plays here the role of the test function. Here $\delta^\varepsilon(\rr)$ is a non-singular Dirac-delta sequence with parametric dependence. For $\varepsilon$ finite this corresponds to considering a line source with rotationally-invariant core of radius $\varepsilon$. We start with the regularization of the dislocation density and dislocation current tensors. The regularization of the dislocation density tensor of a Volterra dislocation is denoted by the convolution product
\begin{align}
\label{alpha-reg}
\alpha^{\text{iso}}_{ij}(\rr,t)=[\alpha_{ij}*\delta^\varepsilon](\rr,t)
=\int\alpha_{ij}(\rr-\rr',t)\, \delta^\varepsilon(\rr')\, \d \rr'
\,,
\end{align}
where $*$ denotes the two-dimensional spatial convolution. The regularized dislocation density tensor reads
\begin{align}
\label{DD-reg}
\alpha^{\text{iso}}_{ij}=\frac{b_i\ell_j}{2\pi\varepsilon^2}\,
\frac{1}{\big[\big(R(t)/\varepsilon\big)^2+1\big]^{3/2}}\,.
\end{align}
The dislocation density tensor $\alpha^{\text{iso}}_{ij}$ is finite and reaches its maximum value of $b_i\ell_j/(2\pi\varepsilon^2)$ at the dislocation core center (Fig.~\ref{fig:alpha}).
Therefore, $\delta^\varepsilon(\rr)$ plays the role of the dislocation shape function in the regularization.

To further motivate this somewhat ad-hoc regularization, it is interesting to compare it with one of a more fundamental nature. Thus, Fig.~\ref{fig:alpha} also displays the dislocation density tensor, obtained in gradient elasticity of Helmholtz type~\citep{LMA05,LM06,Lazar14},
\begin{align}
\label{DD-grad}
\alpha_{ij}=\frac{b_i\ell_j}{2\pi\varepsilon^2}\, K_0 \big(R(t)/\varepsilon\big),
\end{align}
which has a weak (logarithmic) residual singularity at the dislocation line. Gradient elasticity of the Helmholtz type serves a regularization based on higher order partial differential equations where the corresponding regularization function is the Green function of the Helmholtz operator~\citep{Lazar14}.
In the intermediate range, the dislocation density tensors~(\ref{DD-reg}) and (\ref{DD-grad}) are in surprisingly good agreement (see Fig.~\ref{fig:alpha}) in spite of markedly different asymptotic behaviors.\footnote{The modified Bessel function behaves as $K_0(x)\sim \sqrt{\pi/2}\,x^{-1/2}{\rm e}^{-x}$ when $x\gg 1$ and $K_0(x)\sim -\ln (x/2) -C$ for $0<x\le1$ ($C$ denotes  Euler's constant).} Moreover, the function~(\ref{DD-reg}) is finite everywhere, in contrast to (\ref{DD-grad}).

The regularized dislocation current tensor is given by
\begin{align}
\label{I-reg}
I^{\text{iso}}_{ij}(\rr,t)=[I_{ij}*\delta^\varepsilon](\rr,t)\,.
\end{align}
The regularized elastic distortion tensor and  elastic velocity vector are defined, respectively, by
\begin{subequations}
\begin{align}
\label{beta-reg}
\beta^{\text{iso}}_{ij}(\rr,t)=[\beta_{ij}*\delta^\varepsilon](\rr,t)\,,\\
\label{v-reg}
v^{\text{iso}}_{i}(\rr,t)=[v_{i}*\delta^\varepsilon](\rr,t)\,.
\end{align}
\end{subequations}
Using the property of the differentiation of a convolution (see, e.g.,~\citet{Vl})
and the equations of motion for the elastic fields~(\ref{B-NE}) and (\ref{v-NE}),
we can show that the regularized elastic fields~(\ref{beta-reg}) and (\ref{v-reg})
satisfy the following inhomogeneous Navier equations
\begin{subequations}
\begin{align}
\label{B-NE-reg}
&L_{ik} \beta^{\text{iso}}_{km}=
L_{ik} [\beta_{km}*\delta^\varepsilon]=
[L_{ik} \beta_{km}]*\delta^\varepsilon=
[\epsilon_{nml}C_{ijkl}\alpha_{kn,j}+\rho\,\dot{I}_{im}]*\delta^\varepsilon
=\epsilon_{nml}C_{ijkl}\alpha^{\text{iso}}_{kn,j}+\rho\,\dot{I}^{\text{iso}}_{im}
\, ,\\
\label{v-NE-reg}
&L_{ik} v^{\text{iso}}_{k}=L_{ik} [v_{k}*\delta^\varepsilon]=
[L_{ik} v_{k}]*\delta^\varepsilon
=[C_{ijkl}\,I_{kl,j}]*\delta^\varepsilon
=C_{ijkl}\,I^{\text{iso}}_{kl,j}\, ,
\end{align}
\end{subequations}
with the regularized dislocation density tensor~(\ref{alpha-reg}) and the
regularized dislocation current tensor~(\ref{I-reg}) as inhomogeneous parts.
In addition, using Eqs.~(\ref{BI1}) and (\ref{BI2}), it can be shown that the
regularized dislocation density tensor~(\ref{alpha-reg}) and the regularized
dislocation current tensor~(\ref{I-reg}) satisfy Bianchi identities
\begin{subequations}
\begin{align}
\label{BI1-red}
\alpha^{\text{iso}}_{ij,j}&=\pd_j[\alpha_{ij}*\delta^\varepsilon]
=[\alpha_{ij,j}]*\delta^\varepsilon=0\, ,\\
\label{BI2-red}
\dot{\alpha}^{\text{iso}}_{ij} &=\pd_t[\alpha_{ij}*\delta^\varepsilon]
=[\dot{\alpha}_{ij}]*\delta^\varepsilon
=-[\epsilon_{jkl}I_{ik,l}]*\delta^\varepsilon
= - \epsilon_{jkl}I^{\text{iso}}_{ik,l}\, .
\end{align}
\end{subequations}

\subsection{Regularized fields}
\label{sec4-2}
The regularized elastodynamic fields are now derived using the isotropic regularization. Eq.\ (\ref{dd-e}) is not specific to edge dislocations, but applies to screw dislocations as well upon taking $\ell_i=\delta_{iz}$ and $b_i=b_z\delta_{iz}$. We therefore start from that expression. Using the isotropic-regularized form of $\alpha_{ij}$
\begin{align}
\alpha^{\rm iso}_{ij}=b_i\ell_j\delta^\varepsilon(\BR(t)),
\end{align}
substituting into Eqs.\ (\ref{B-1}) and (\ref{v-1}), and performing the $\rr'$-integration we obtain the regularized fields in the form
\begin{subequations}
\label{eq:isoints}
\begin{align}
\label{betaiso}
\beta^{\rm iso}_{ij}(\rr,t)&=
\epsilon_{njl}b_p\ell_n\int_{-\infty}^{t^-} \Big\{G^{\rm iso}_{iq,k}(\rr-\Bs(t'), t-t')\widetilde{C}_{kqpl}\big(\BV(t')\big)
+\rho\, G^{\rm iso}_{ip}(\rr-\Bs(t'), t-t') \dot{V}_l(t')\Big\}\,\d t'\,,\\
\label{viso}
v_i^{\rm iso}(\rr,t)&=\epsilon_{nml}b_{p}\ell_nC_{jkpm}\int_{-\infty}^{t^-} G^{\rm iso}_{ij,k}(\rr-\Bs(t'), t-t') V_l(t')\,\d t'\,,
\end{align}
\end{subequations}
where the regularized Green tensor function is
\begin{align}
\label{GT-reg-conv}
G^{\text{iso}}_{ij}(\rr,t)=[G^+_{ij}*\delta^\varepsilon](\rr,t)=
\int G^+_{ij}(\rr-\rr',t)\,\delta^\varepsilon(\rr')\, \d \rr',
\end{align}
and where the upper boundary has been chosen as $t^-$ (slightly less than $t$), according to the remark made in Sec.\ \ref{sec:remarks}. The latter convention is used throughout the rest of the paper.

The regularized Green tensor satisfies the inhomogeneous Navier equation
\begin{align}
\label{GF-e-reg}
L_{ik} G^{\text{iso}}_{km}(\rr-\rr',t-t')=\delta_{im}\,
\delta(t-t')\,\delta^\varepsilon(\rr-\rr')
\, .
\end{align}
Carrying out the convolution in Eq.\ (\ref{GT-reg-conv}) yields the remarkable result that due to our choice for $\delta^\varepsilon$ the regularized form $G^{\text{iso}}(\rr,t)$ of the \emph{distribution} $G^+_{ij}(\rr,t)$ is conveniently expressed in terms of the \emph{function} $G_{ij}(\rr,t)$, continued to complex time, as (\cite{PL2014})
\begin{subequations}
\label{eq:gisoijzz}
\begin{align}
\label{eq:gisoij}
G^{\text{iso}}_{ij}(\rr,t)&=\theta(t)\Re\left[G_{ij}(\rr,t)_{c t\to c t+\ii\varepsilon}\right]\qquad(i,j=x,y),\\
\label{eq:gisozz}
G^{\text{iso}}_{zz}(\rr,t)&=\theta(t)\Re\left[G_{zz}(\rr,t)_{c_{\rm T} t\to c_{\rm T} t+\ii\varepsilon}\right],
\end{align}
\end{subequations}
where our notations mean that $c_{\rm T} t$ and $c_{\rm L} t$ must be replaced in $G_{ij}(\rr,t)$ by $c_{\rm T} t+\ii\varepsilon$ and $c_{\rm L}t+\ii\varepsilon$, respectively, according to the remark following Eq.\ (\ref{GT-2D-grad}).

The function $G_{ij}(\rr,t)$ is readily deduced from the associated distribution $G_{ij}^+(\rr,t)$ by removing causality and wavefront constraints on its variables (i.e., in practice, by simply removing the $\theta(t)$ prefactor and the `plus' subscripts), which allows for its continuation to complex-valued arguments. For instance, in the antiplane-strain case,
\begin{align}
\label{eq:Gzzfunc}
G_{zz}(\rr,t)=\frac{1}{2\pi\mu}\left(t^2-r^2/\cT^2\right)^{-1/2},
\end{align}
to be compared with (\ref{GT-zz}). The function $G_{ij}(\rr,t)$ of the plane-strain case is obtained from Eq.\ (\ref{GT-2D}) in the same manner. Similarly, the regularization of the gradient of the Green tensor is given by
\begin{align}
\label{eq:Gisoijk}
G^{\text{iso}}_{ij,k}(\rr,t)=\theta(t)\Re \left[G_{ij,k}(\rr,t)_{c t\to c t+\ii\varepsilon}\right]
\end{align}
where $G_{ij,k}(\rr,t)$ is the function that can be read from the distributional expressions of the gradients (\ref{GT-zz-grad}) (anti-plane-strain) or (\ref{GT-2D-grad}) (plane-strain), removing as above $\theta(t)$, the `plus' subscripts, and the `Pf' prescriptions. Analytic continuation of functions has long been known as a method of representing pseudofunctions~\citep{BREM61,GS}. Indeed, upon taking the limit $\varepsilon\to 0^+$ Eqs.\ (\ref{eq:gisoijzz}) and (\ref{eq:Gisoijk}) induce definitions of the distributions $G^+_{ij}$ and $G^+_{ij,k}$ as
\begin{align}
\label{eq:limitdistr}
G_{ij}^+(\rr,t)=\lim_{\varepsilon\to 0^+}G^{\text{iso}}_{ij}(\rr,t),\qquad G_{ij,k}^+(\rr,t)=\lim_{\varepsilon\to 0^+}G^{\text{iso}}_{ij,k}(\rr,t).
\end{align}

The functions $G^{\text{iso}}_{ij}(\rr,t)$ and $G^{\text{iso}}_{ij,k}(\rr,t)$ are nowhere singular in the $\rr$-plane, and possess equal-time limits similar to Eq.\ (\ref{G-et-lim}) (Appendix A):
\begin{align}
\label{eq:equal-time-iso}
\lim_{t\to 0^+}G^{\text{iso}}_{ij}(\rr,t)=0,\qquad \lim_{t\to 0^+}\partial_t G^{\text{iso}}_{ij}(\rr,t)=\rho^{-1}\delta_{ij}\delta^\varepsilon(\rr).
\end{align}

For uniform motion $\BV(t)\equiv\BV$ and $\Bs(t)=\BV t$. Then, letting $\tau=t-t'$,  the regularized field expressions (\ref{betaiso}) and (\ref{viso}) reduce to
\begin{subequations}
\label{eq:isounif}
\begin{align}
\label{betaisounif}
\beta^{\rm iso}_{ij}(\rr,t)&=
\epsilon_{njl}b_p\ell_n\widetilde{C}_{kqlp}(\BV)\int_{0^+}^{+\infty}G^{\rm iso}_{iq,k}(\rr-\BV t+\BV\tau,\tau)\,\d \tau\,,\\
\label{visounif}
v_i^{\rm iso}(\rr,t)&=\epsilon_{nml}b_{p}\ell_n C_{jkmp}V_l\int_{0^+}^{+\infty}G^{\rm iso}_{ij,k}(\rr-\BV t+\BV\tau,\tau)\,\d \tau\,.
\end{align}
\end{subequations}
Such steady-state fields are usually computed in the co-moving frame centered on the dislocation. This change of origin, which consists in turning the position vector $\br$ into $\BV t+\rr$, removes the trivial time dependence in (\ref{betaisounif}) and (\ref{visounif}). In particular, the static fields  ($\BV=\mathbf{0}$) read
\begin{subequations}
\label{eq:isostat0}
\begin{align}
\label{betaisotat0}
\beta^{\rm iso}_{ij}(\rr,t)&=\epsilon_{njl}b_p\ell_n C_{kqlp} \int_{0^+}^{+\infty}G^{\rm iso}_{iq,k}(\rr,\tau)\,\d \tau\,,\\
\label{visostat0}
v_i^{\rm iso}(\rr,t)&=0\,.
\end{align}
\end{subequations}
Due to the symmetries in the indices $l$ and $m$, one can replace $C_{jkmp}$ by $\widetilde{C}_{jkmp}$ in Eq.~(\ref{visounif}). Thus,
\begin{align}
\label{visounif2}
v_i^{\rm iso}(\rr,t)&=\epsilon_{nml}b_{p}\ell_n\widetilde{C}_{jkmp}(\BV)V_l\int_{0^+}^{+\infty}G^{\rm iso}_{ij,k}(\rr-\BV t+\BV\tau,\tau)\,\d \tau\,.
\end{align}
We thus retrieve the following relation for uniform motion between the elastic velocity and the elastic distortion, which is as a direct consequence of the equation $v_i=\dot{u}_i$:
\begin{align}
\label{rel-VB}
v_i^{\rm iso}=-V_j \beta^{\rm iso}_{ij}\qquad\text{(uniform motion)}\,.
\end{align}

We focus hereafter on non-uniform motions that begin at $t=0$, starting from a steady state of constant initial velocity $\BV^{(0)}$ at times $t<0$. The contributions of negative times can then be separated out into the following integral, which differs from the ones in Eqs.\ (\ref{eq:isounif}) by the lower integration bound:
\begin{align}
\label{eq:iiso0}
I^{{\rm iso}(0)}_{ijk}(\rr,t)&=\int_{-\infty}^{0^-} G^{\rm iso}_{ij,k}(\rr-\BV^{(0)}t', t-t')\, \d t'=\int_{t^+}^{+\infty}G^{\rm iso}_{ij,k}(\rr-\BV^{(0)}t+\BV^{(0)}\tau,\tau)\,\d\tau\,.
\end{align}
Then, Eqs.\ (\ref{eq:isoints}) read, for $t>0$,
\begin{subequations}
\label{iso1}
\begin{align}
\label{Biso1}
\beta^{\rm iso}_{ij}(\rr,t)&=\epsilon_{njl}b_p\ell_n\biggl\{I^{{\rm iso}(0)}_{iqk}(\rr,t)\widetilde{C}_{kqpl}(\BV^{(0)})\\
&{}+\int_{0^-}^{t^-} \Big[G^{\rm iso}_{iq,k}(\rr-\Bs(t'), t-t')\widetilde{C}_{kqpl}\big(\BV(t')\big)
+\rho\, G^{\rm iso}_{ip}(\rr-\Bs(t'), t-t') \dot{V}_l(t')\Big]\d t'
\biggr\},\nonumber\\
\label{viso1}
v_i^{\rm iso}(\rr,t)&=\epsilon_{nml}b_{p}\ell_nC_{jkpm}\left[I^{{\rm iso}(0)}_{ijk}(\rr,t)V_l^{(0)}+\int_{0^-}^{t^-}G^{\rm iso}_{ij,k}(\rr-\Bs(t'), t-t') V_l(t')\,\d t'\right]\,,
\end{align}
\end{subequations}
whereas at negative times the fields are given by Eqs.\ (\ref{eq:isounif}) with $\BV=\BV^{(0)}$. Integral (\ref{eq:iiso0}) vanishes as $t\to+\infty$, accounting for the `afterglow-type' progressive erasure of the steady-state field that was present prior to non-uniform motion \citep{PELL14}.

\section{Implementation}
\label{sec5}
We now examine a way of handling Eqs.\ (\ref{iso1}) for numerical purposes.
\subsection{Discrete representation of motion}
Following a series of studies devoted to the study of inertial effects during non-uniform dislocation motion \citep{PILL07,PILL09,PELL14}, our discretization scheme consists in transforming the physical velocity function $\BV(t)$ into a piecewise-constant function, whose constant-valued pieces are separated by a finite number of velocity jumps. Specifically, motion is split into $N(t)+1$ time intervals $]t_{\gamma-1},t_\gamma[$, $0\leq \gamma\leq N$ of constant velocity $\BV^{(\gamma)}$. By convention the first interval $\gamma=0$ is the semi-infinite one of negative times, with $t_{-1}=-\infty$ and $t_0=0^-$. Also, the last interval $\gamma=N$ is conventionally bounded upwards by the current time, so that $t_N=t^-$. The other ones are of arbitrary duration. The integer $N(t)$ represents the number of velocity jumps that have occurred up to time $t$. The velocity jumps are $\Delta\BV^{(\gamma)}=\BV^{(\gamma)}-\BV^{(\gamma-1)}$. The velocity and acceleration are thus represented as
\begin{subequations}
\begin{align}
\label{eq:vel}
\BV(t)&=\BV^{(N(t))}=\BV^{(0)}+\sum_{\gamma=1}^{N(t)}\theta(t-t_{\gamma-1})\Delta\BV^{(\gamma)},\\
\label{eq:acc}
\dot{\BV}(t)&=\sum_{\gamma=1}^{N(t)}\delta(t-t_{\gamma-1})\Delta\BV^{(\gamma)}.
\end{align}
\end{subequations}
Introducing discrete positions at jump times
\begin{align}
\label{eq:sgamdef}
\Bs_\gamma=\sum_{\gamma'=1}^{\gamma}(t_{\gamma'}-t_{\gamma'-1})\BV^{(\gamma')}\qquad(\gamma<N),
\end{align}
the position reads, consistently with (\ref{eq:vel}),
\begin{align}
\label{eq:stsk}
\Bs(t)=\left\{
\begin{array}{ll}
\BV^{(0)}t &\qquad\text{if }t<0\\
\Bs_{N-1}+(t-t_{N-1})\BV^{(N)}&\qquad\text{if }t>0,
\end{array}
\right..
\end{align}

\subsection{Fields as sums of closed-form time integrals}
\label{sec:sec5-2}
Expanding the time integrals (\ref{iso1}) on the set of constant-velocity intervals and using (\ref{eq:acc}) yields
\begin{subequations}
\label{eq:iso2}
\begin{align}
\beta^{\rm iso}_{ij}(\rr,t)&=
\epsilon_{njl}b_p\ell_n
\biggl[I^{{\rm iso}(0)}_{iqk}(\rr,t)\widetilde{C}_{kqpl}\big(\BV^{(0)}\big)+\sum_{\gamma=1}^{N(t)}
\widetilde{C}_{kqpl}\big(\BV^{(\gamma)}\big)\int_{t_{\gamma-1}}^{t_\gamma} G^{\rm iso}_{iq,k}(\rr-\Bs(t'), t-t')\,\d t'
\nonumber\\
\label{Biso2}
&\hspace{4.3cm}{}+\rho\sum_{\gamma=1}^{N(t)} G^{\rm iso}_{ip}(\rr-\Bs_{\gamma-1}, t-t_{\gamma-1}) \Delta V_l^{(\gamma)}\biggr]\,,\\
\label{viso2}
v_i^{\rm iso}(\rr,t)&=
\epsilon_{nml}b_{p}\ell_nC_{jkpm}\left[I^{{\rm iso}(0)}_{ijk}(\rr,t)V^{(0)}_l+\sum_{\gamma=1}^{N(t)}V^{(\gamma)}_l\int_{t_{\gamma-1}}^{t_\gamma} G^{\rm iso}_{ij,k}(\rr-\Bs(t'), t-t')\, \d t'\right].
\end{align}
\end{subequations}
The most important building-block of Eqs.\ (\ref{eq:iso2}) is the time integral
\begin{align}
\label{eq:igam0}
I^{{\rm iso}(\gamma)}_{ijk}(\rr,t)=\int_{t_{\gamma-1}}^{t_\gamma} G^{\rm iso}_{ij,k}(\rr-\Bs(t'), t-t')\, \d t'\,,
\end{align}
which generalizes (\ref{eq:iiso0}). Rewriting it by means of (\ref{eq:stsk}), it reduces to
\begin{align}
\label{eq:igam1}
I^{{\rm iso}(\gamma)}_{ijk}(\rr,t)
&=\int_{t_{\gamma-1}}^{t_\gamma}G^{\rm iso}_{ij,k}(\rr-[\Bs_{\gamma-1}+(t-t_{\gamma-1})\BV^{(\gamma)}]+(t-t')\BV^{(\gamma)}, t-t')\, \d t'\,,
\end{align}
where an extra term $t\BV^{(\gamma)}$ has been added and subtracted in the first slot of $G^{\rm iso}_{ij,k}$. The new vector involved,
\begin{align}
\label{eq:svirtdef}
\Bs^{\rm virt}_\gamma(t)=\Bs_{\gamma-1}+(t-t_{\gamma-1})\BV^{(\gamma)},
\end{align}
represents the \emph{virtual} position that the dislocation would have as instant $t$ if motion had continued at uniform velocity $\BV^{(\gamma)}$ after the velocity jump at $t_{\gamma-1}$. Such virtual motions determine fields in remote regions of space that have not yet been swept by subsequent acceleration waves \citep{PILL09}. Introducing the \emph{indefinite} integral
\begin{align}
\label{eq:jiso}
J^{\rm iso}_{ijk}(\rr,t;\BV)&=\int^{t}G^{\rm  iso}_{ij,k}(\rr+\tau\BV,\tau)\,\d\tau\, ,
\end{align}
and letting again $\tau=t-t'$, integral (\ref{eq:igam1}) now reads
\begin{subequations}
\begin{align}
I^{{\rm iso}(\gamma)}_{ijk}(\rr,t)
&=\int^{t-t_{\gamma-1}}_{t-t_\gamma}G^{\rm iso}_{ij,k}(\rr-\Bs^{\rm virt}_\gamma(t)+\tau\BV^{(\gamma)},\tau)\, \d \tau\nonumber\\
\label{eq:igam2}
&=J^{\rm iso}_{ijk}(\rr-\Bs^{\rm virt}_\gamma(t),t-t_{\gamma-1};\BV^{(\gamma)})-J^{\rm iso}_{ijk}(\rr-\Bs^{\rm virt}_\gamma(t),t-t_\gamma;\BV^{(\gamma)})\,.
\end{align}
In particular, because $t_N=t^-$, the last term $\gamma=N$ reads
\begin{align}
\label{eq:iisoN}
I^{{\rm iso}(N)}_{ijk}(\rr,t)
&=J^{\rm iso}_{ijk}(\rr-\Bs^{\rm virt}_\gamma(t),t-t_{N-1};\BV^{(\gamma)})-J^{\rm iso}_{ijk}(\rr-\Bs^{\rm virt}_\gamma(t),0^+;\BV^{(\gamma)})\,.
\end{align}
An equation analogous to (\ref{eq:igam2}) applies as well to the $\gamma=0$ term, since by (\ref{eq:iiso0}) and (\ref{eq:jiso}),
\begin{align}
\label{eq:iiso0J}
I^{{\rm iso}(0)}_{ijk}(\rr,t)&=J^{\rm iso}_{ijk}(\rr-\BV^{(0)}t,+\infty;\BV^{(0)})-J^{\rm iso}_{ijk}(\rr-\BV^{(0)}t,t^+;\BV^{(0)})\,.
\end{align}
\end{subequations}

Closed-form expressions for the function $J^{\rm iso}_{ijk}(\rr,t;\BV)$, derived from the latter reference, are summarized in Appendix \ref{sec:appA} [Eqs.\ (\ref{eq:calJzzk}), (\ref{eq:Jisozzk}) and (\ref{eq:calJijk}), (\ref{eq:Jisoijk})]. Closed-form expressions are provided as well for the limiting functions $J^{\rm iso}_{ijk}(\rr,+\infty;\BV)$, needed in (\ref{eq:iiso0J}) [Eqs.\ (\ref{eq:Jisozzk}), (\ref{eq:limitauinfzzk}) and (\ref{eq:Jisoijk}), (\ref{eq:limitauinfijk})]. It is shown in Appendix \ref{sec:doublelim} that the following limits commute:
\begin{align}
\lim_{\tau\to+\infty}\lim_{V\to 0}J^{\rm iso}_{ijk}(\rr,\tau;\BV)=\lim_{V\to
  0}\lim_{\tau\to+\infty}J^{\rm iso}_{ijk}(\rr,\tau;\BV)\, ,
\end{align}
so that the static fields at $\BV=0$ are well-defined.

Therefore, the fields (\ref{eq:iso2}) finally take the following form, to be used in numerical computations:
\begin{subequations}
\label{eq:iso3}
\begin{align}
\label{Biso3}
\beta^{\rm iso}_{ij}(\rr,t)&=\epsilon_{njl}b_p\ell_n\biggl[\sum_{\gamma=0}^{N(t)}I^{{\rm iso}(\gamma)}_{iqk}(\rr,t)\widetilde{C}_{kqpl}\big(\BV^{(\gamma)}\big)
+\rho\sum_{\gamma=1}^{N(t)} G^{\rm iso}_{ip}(\rr-\Bs_{\gamma-1}, t-t_{\gamma-1}) \Delta V_l^{(\gamma)}\biggr]\,,\\
\label{viso3}
v_i^{\rm iso}(\rr,t)
&=\epsilon_{nml}b_{p}\ell_nC_{jkpm}\sum_{\gamma=0}^{N(t)}V^{(\gamma)}_l I^{{\rm iso}(\gamma)}_{ijk}(\rr,t)\,.
\end{align}
\end{subequations}

The writing (\ref{eq:igam2}) of the definite integral as a difference of boundary values of the indefinite integral (\ref{eq:jiso}) is the key step of the computational procedure. It requires the integrand to be analytic in the immediate vicinity of the integration intervals (integration paths). This is warranted by the isotropic regularization employed, which ensures that no branch cut of $G^{\rm iso}_{ij,k}(\rr+\tau\BV,\tau)$ is crossed as $\tau$ varies within these intervals (\citet{PL2014}).

Accordingly, in Eq.\ (\ref{eq:iisoN}) resides the ultimate justification of our using $t^-$ as an upper boundary in the time integrals of Eqs.\ (\ref{eq:isounif}): indeed, $t'=t$ is a point of analyticity breakdown beyond which the Green tensor and its gradient vanish identically by causality, so that using either $t$ or $t^+$ does not allow one to employ the integration formula (\ref{eq:igam2}), contrary to using $t^-$.

The above expressions have been implemented in a Fortran code, employed to produce the field maps below, in which the prescriptions $t^+$ and $0^+$ are translated as $t+\eta$ and $\eta$, with $\eta=10^{-5}$.

\subsection{Steady fields (uniform motion)}
\label{sec:steadyfields}
The particular case of uniform motion is addressed by letting $\BV^{(\gamma)}\equiv\BV$ for all $\gamma$, so that $\Delta\BV^{(\gamma)}\equiv 0$. Equations (\ref{eq:iso3}) simplify as
\begin{subequations}
\label{eq:iso3steady}
\begin{align}
\label{Biso3steady}
\beta^{\rm iso}_{ij}(\rr,t)
&=\epsilon_{njl}b_p\ell_n\widetilde{C}_{kqpl}\big(\BV\big)\sum_{\gamma=0}^{N(t)}I^{{\rm iso}(\gamma)}_{iqk}(\rr,t)\,,\\
\label{viso3steady}
v_i^{\rm iso}(\rr,t)
&=\epsilon_{nml}b_{p}\ell_nC_{jkpm}V_l\sum_{\gamma=0}^{N(t)} I^{{\rm iso}(\gamma)}_{ijk}(\rr,t)\,.
\end{align}
\end{subequations}
Moreover, by (\ref{eq:sgamdef}), one has
\begin{align}
\Bs_\gamma=\BV\sum_{\gamma'=1}^{\gamma}(t_{\gamma'}-t_{\gamma'-1})=\BV(t_\gamma-t_0)=t_\gamma\BV.
\end{align}
The virtual positions (\ref{eq:svirtdef}) then reduce to $\Bs^{\rm virt}_\gamma(t)\equiv \bV t$ for all $\gamma$. Using the fact that $t_0=0^-$, and expressions (\ref{eq:igam2}), (\ref{eq:iisoN}), and (\ref{eq:iiso0J}), it follows that
\begin{align}
\sum_{\gamma=0}^{N(t)} I^{{\rm iso}(\gamma)}_{ijk}(\rr,t)=J^{\rm iso}_{ijk}(\rr-\BV t,+\infty;\BV)-J^{\rm iso}_{ijk}(\rr-\BV t,0^+;\BV)\,.
\end{align}
Substituting the latter expression into Eqs.\ (\ref{eq:iso3steady}), we deduce that in the co-moving frame the fields are time-independent, and read
\begin{subequations}
\label{eq:Bvisosteady}
\begin{align}
\label{BisoSteady}
\beta^{\rm iso}_{ij}(\rr)
&=\epsilon_{njl}b_p\ell_n\widetilde{C}_{kqpl}\big(\BV\big)\left[J^{\rm iso}_{iqk}(\rr,+\infty;\BV)-J^{\rm iso}_{iqk}(\rr,0^+;\BV)\right]
\qquad\text{(co-moving frame)}\,,\\
\label{visoSteady}
v_i^{\rm iso}(\rr)
&=\epsilon_{nml}b_{p}\ell_nC_{jkpm}V_l \left[J^{\rm iso}_{ijk}(\rr,+\infty;\BV)-J^{\rm iso}_{ijk}(\rr,0^+;\BV)\right]
\qquad\text{(co-moving frame)}\,.
\end{align}
\end{subequations}
Classical (i.e., non-distributional) expressions for the (singular) fields of a uniformly-moving Volterra dislocation, valid for velocities $|V|<\cT$ can be retrieved by abruptly setting $\varepsilon=0$ in those expressions (no limit process). In this case, the second term cancels out (see Appendix \ref{sec:valtzer}), so that
\begin{subequations}
\label{eq:BvVoltSteady}
\begin{align}
\label{BVoltSteady}
\beta^{\rm Volterra}_{ij}(\rr)
&=\epsilon_{njl}b_p\ell_n\widetilde{C}_{kqpl}\big(\BV\big) J^{\rm iso}_{iqk}(\rr,+\infty;\BV)|_{\varepsilon=0}\qquad\text{(co-moving frame)}\,,\\
\label{vVoltSteady}
v_i^{\rm Volterra}(\rr)
&=\epsilon_{nml}b_{p}\ell_n C_{jkpm} V_l J^{\rm iso}_{ijk}(\rr,+\infty;\BV)|_{\varepsilon=0}\qquad\text{(co-moving frame)}\,.
\end{align}
\end{subequations}
The main difference between both sets of expressions is that due to the finite core size $\varepsilon$, fields given by Eq.~(\ref{eq:Bvisosteady}) are non-singular everywhere, and correctly display one ($\cT<|V|<\cL$) or two ($|V|>\cL$) Mach cones for faster-than-wave velocities. Moreover, Volterra-dislocation fields can be made to exhibit Mach cones only by carefully taking the limit $\varepsilon\to 0$ in the sense of distributions \citep{PL2014}. However, the latter cones are supported by infinitely thin Dirac lines, and thus cannot be rendered in field maps. In general, distributional expressions are not directly suitable to full-field representation.

\begin{figure}
\begin{centering}
\includegraphics[width=16cm]{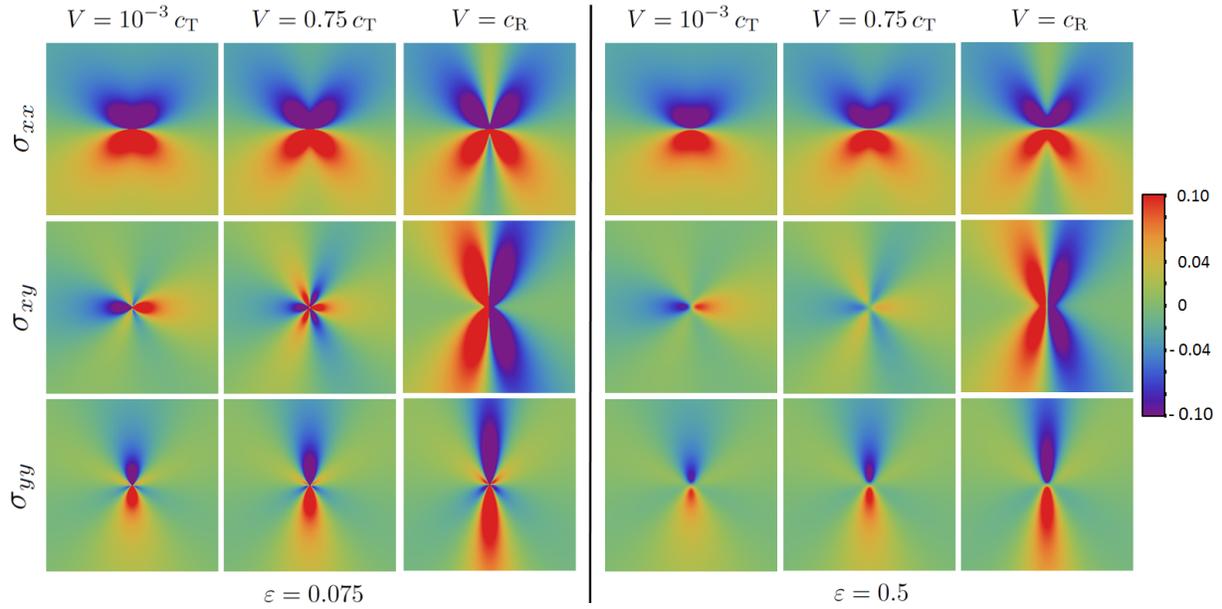}
\caption{\label{fig:fig2}
Effect of the regularizing core width $\varepsilon$ on the stress field components of a `glide' edge dislocation, in steady horizontal motion in the positive direction, at low velocities $V<c_{\rm R}$. Velocity and $\varepsilon$ as indicated. For better display, stress levels have been thresholded as indicated in the bar legend.}
\end{centering}
\end{figure}
Figure \ref{fig:fig2} displays the regularized stress field components of a `glide' edge dislocation ($\Bb=(1,0,0)$) in steady horizontal motion, computed in the co-moving frame from Eqs.\ (\ref{eq:Bvisosteady}) for some velocities $V$ less than the Rayleigh velocity $c_{\rm R}$ \citep{WW80,HIRT82}, and for two regularizing core widths $\varepsilon$. Material constants are such that the wavespeed ratio is $\cL/\cT=2.2$ ($\lambda=2.84$), so that $c_{\rm R}\simeq 0.937096\,\cT$. Units are taken dimensionless, such that $\cT=1$ and $\mu=1$, and the box size is $L_x\times L_y=10\times 10$. For $\varepsilon=0.075$, the new regularized fields, albeit smooth, are very close to the standard results for a Volterra dislocation [see Eqs.\ (7-24) to (7-26) in \citep{HIRT82}]\footnote{Actually, the overall sign of the stress components for $V\not =0$ given in Hirth and Lothe's treatise must be changed to match our own expressions, as well as the standard static ones in Eq.\ (3-43) of the same reference in the limit $V\to 0$.} For $V=10^{-3}\cT$, the fields are nearly identical to the static ones. The effect of increasing the core width (to $\varepsilon=0.5$ in the figure) is to reduce the overall stress strength, and to widen the gaps between the different lobes of the field patterns. In those gaps the fields have near-zero values (especially close to the core center), which illustrates the regularizing property of $\varepsilon$.

\begin{figure}
\begin{centering}
\includegraphics[width=16cm]{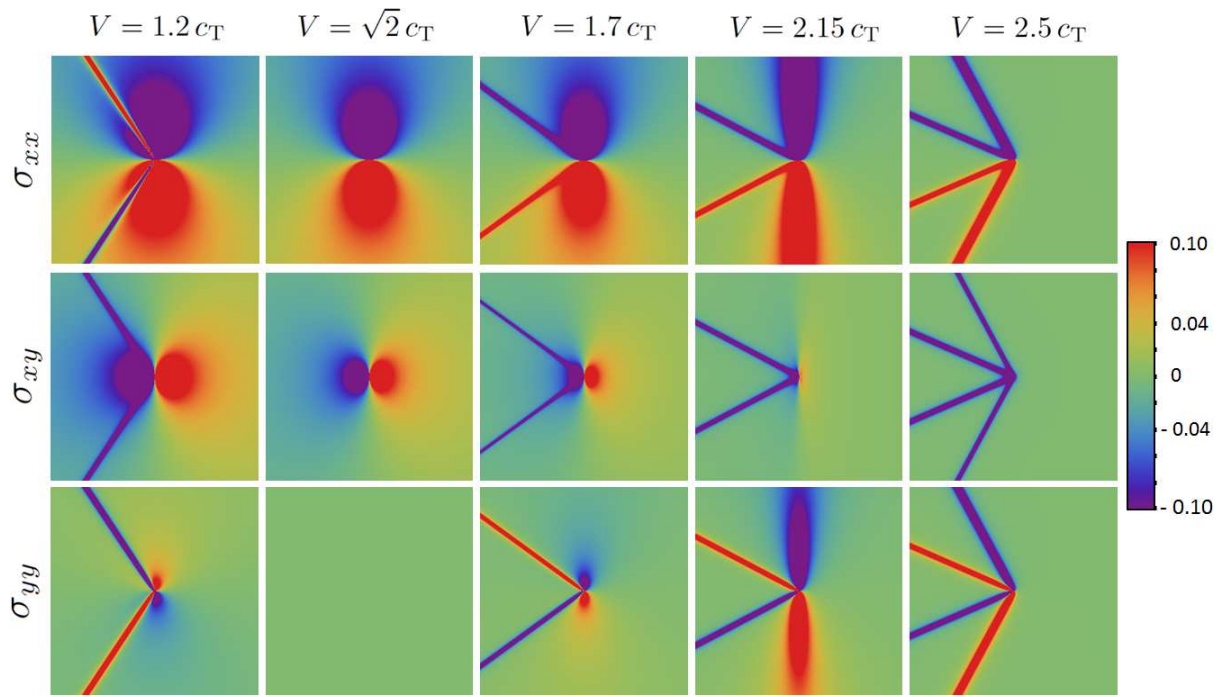}
\caption{\label{fig:fig3}
Stress field components of a `glide' edge dislocation, in faster-than-wave steady horizontal motion in the positive direction, with velocities $V>\cT$, for $\varepsilon=0.075$. Same material parameters and scale as in Fig.\ \ref{fig:fig2}.}
\end{centering}
\end{figure}
Figure \ref{fig:fig3} represents, for $\varepsilon=0.075$ and same material parameters as in Fig.\  \ref{fig:fig2}, the stress components at faster-than wave velocities. In this regime, Mach-cones show up from our analytical field expressions, unlike with classical field expressions. Even in the steady state Mach cones radiate energy to infinity \citep{STRO62}, which gives rise to a finite drag force opposed to uniform dislocation motion \citep{ROSA01}. In the range $\cT<V<\cL$, only the shear-wave Mach cone is present, whereas the lobes are of `longitudinal' character. The shear-wave cone vanishes for the special velocity $V=\sqrt{2}\cT$ at which the dislocation undergoes frictionless motion \citep{ESHE49}. Upon crossing the latter velocity the Mach-cones in $\sigma_{xx}$ and $\sigma_{yy}$ change their sign (single root) whereas the one in $\sigma_{xy}$ vanishes without changing its sign (double root). Since the lobes of $\sigma_{yy}$ change their sign as well, the $\sigma_{yy}$ component vanishes totally at this velocity. Near to the longitudinal wave speed $V\lesssim \cL=2.2\cT$, the lobes of $\sigma_{xx}$ and $\sigma_{yy}$ extend vertically to form an incipient Mach cone associated to the longitudinal wave, whereas the forward (positive) lobe of $\sigma_{xy}$ vanishes. Two pairs of Mach cones make up the field structure at $V>\cL$ with longitudinal branches in the $\sigma_{xx}$ and $\sigma_{yy}$ components of opposite signs (compressive-like above the glide plane, and tensile-like below it), while Mach cones in  $\sigma_{xy}$ are negative. Since in the plane strain set-up the pressure is $p=-(1+\nu)(\sigma_{xx}+\sigma_{yy})/3$, where $\nu=\lambda/[2(\lambda+\mu)]$ is Poisson's ratio, the shear-wave Mach cones of $\sigma_{xx}$ and $\sigma_{yy}$ are of opposite sign and same intensity (see Figure), to cancel out mutually in the pressure field, leaving only a longitudinal Mach cone in pressure for supersonic velocities $V>\cL$ (not shown).

It should be noted that stable steady motion in the `velocity gap' $c_{\rm R}<V<\sqrt{2}\cT$ is impossible on theoretical grounds \citep{ROSA01}. However, as the present work does not not address the equation of motion that drives the dislocation under an external stress, which would forbid such motion, field maps can be computed anyway in this unphysical regime ($V=1.2\,\cT$ in Fig.\ \ref{fig:fig3}).

\section{The two-dimensional elastodynamic Tamm problem for dislocations}
\label{sec6}
In this Section, the formalism is applied to a case of non-uniform motion of physical interest. In electromagnetic field theory, the Tamm problem \citep{TAMM39}, introduced to help elucidating the nature of the Cerenkov radiation, consists in studying the fields radiated by a charge moving in a polarizable medium at faster-than-light velocity during a finite time interval, and at rest otherwise. For a recent review, discussion, and historical account, see \citet{AFAN04}.

We transpose hereafter this problem to the elastodynamic fields radiated by a `glide' edge dislocation (for illustrative purposes, but the method applies to the other two characters as well), using Eqs.\ (\ref{eq:iso3}) with core-width parameter $\varepsilon=0.075$. Material parameters, and dimensionless units, are the same as in the previous Section. A dislocation, initially at rest, is instantaneously accelerated to faster-than-wave speed $V=2.5\cT$ at $t=0$, and moves uniformly at this speed until it is instantaneously pinned at $t=5$ into rest again (e.g., by some impurity or by forest dislocations in intersecting glide planes). Then $N=2$ in Eqs.\ (\ref{eq:iso3}) and the problem allows one to examine fields radiated in both the acceleration and the deceleration steps.

\begin{figure}
\begin{centering}
\includegraphics[width=16cm]{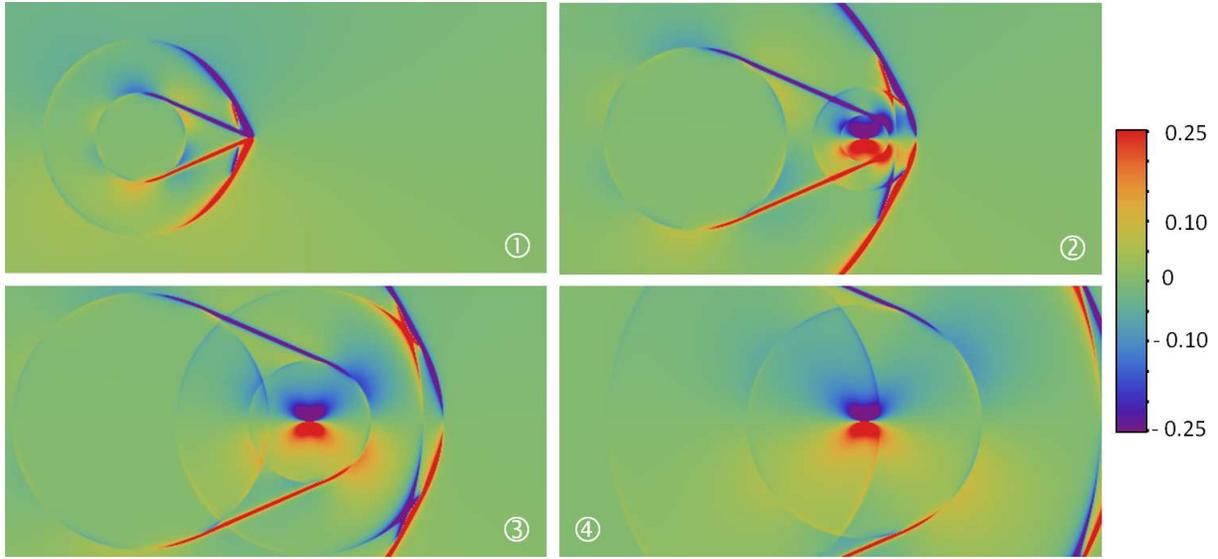}
\caption{\label{fig:fig4}
Stress component $\sigma_{xx}$ of a `glide' edge dislocation in the Tamm problem (see text).}
\end{centering}
\end{figure}

\begin{figure}
\begin{centering}
\includegraphics[width=16cm]{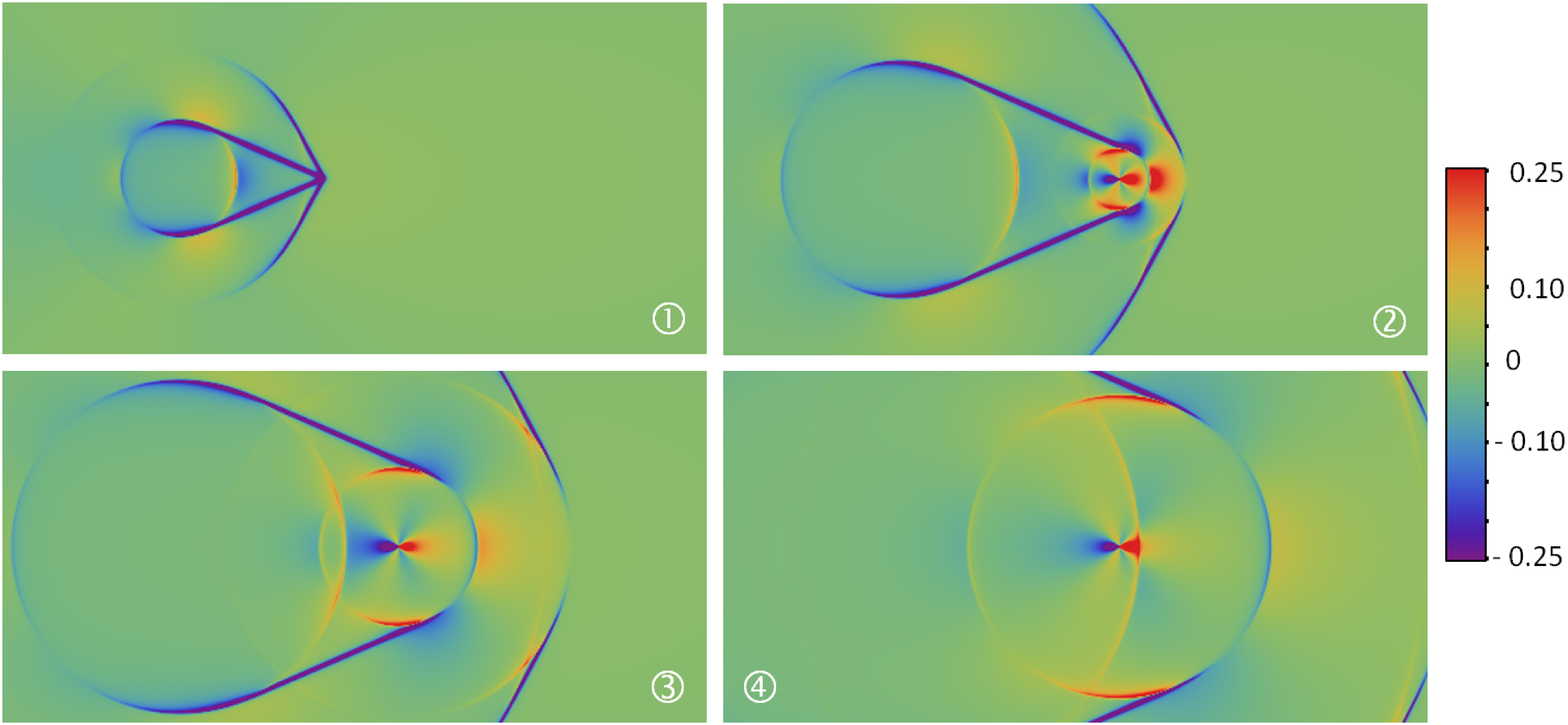}
\caption{\label{fig:fig5}
Stress component $\sigma_{xy}$ of a `glide' edge dislocation in the Tamm problem (see text).}
\end{centering}
\end{figure}

\begin{figure}
\begin{centering}
\includegraphics[width=16cm]{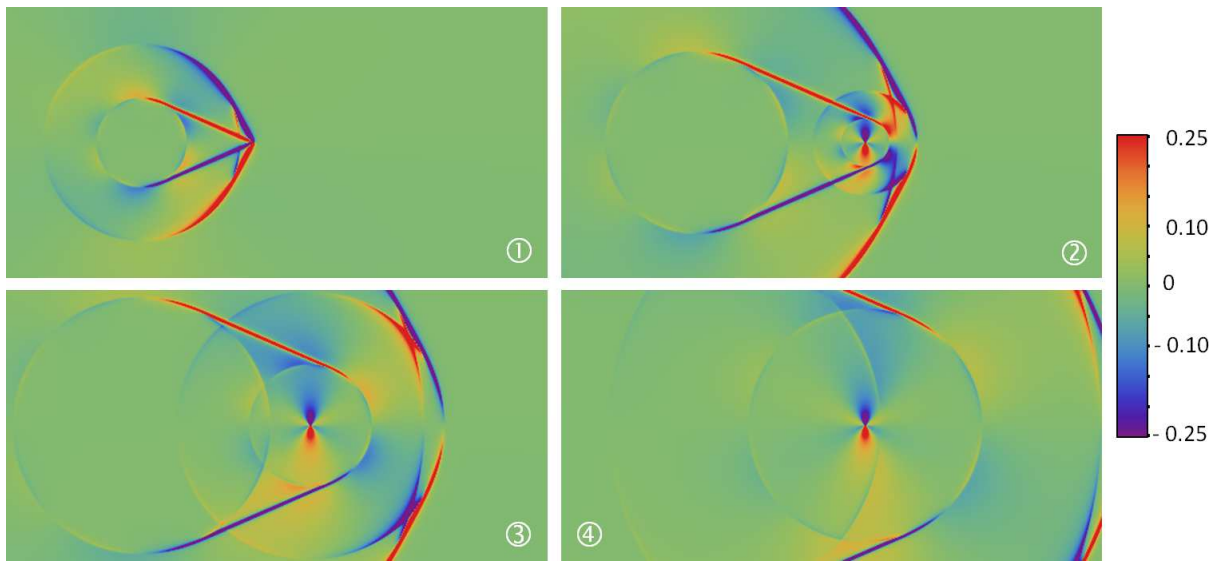}
\caption{\label{fig:fig6}
Stress component $\sigma_{yy}$ of a `glide' edge dislocation in the Tamm problem (see text).}
\end{centering}
\end{figure}

Figs.\ \ref{fig:fig4}, \ref{fig:fig5} and  \ref{fig:fig6} display, respectively, 512$\times$256-pixel pictures of the stress components $\sigma_{xx}$,  $\sigma_{xy}$ and  $\sigma_{yy}$ at times: (1) $t=3.30$; (2), $t=6.74$; (3) $t=9.49$; and (4) $t=13.62$, in a box of physical size $L_x\times L_y=40\times20$. All three components are plotted for further reference. After the initial acceleration, the dislocation velocity is faster than the longitudinal wave, so that two Mach cones build up. In pictures (1), the initial field of the dislocation at rest has already been erased, and the two concentric expanding rings of the acceleration wave, propagating at velocities $\cT$ (inner ring) and $\cL$ (outer ring) control the lateral expansion of the Mach cones. The latter remain tangent to the rings. Images (2) to (4) take place after dislocation sudden pinning, and illustrate the interplay between the acceleration rings, and the braking (Bremsstrahlung) waves. The latter delimit the build-up region of the new static field. After dislocation pinning, the branches of the Mach cones are released to infinity while remaining tangent to the braking rings.  In pictures (2) and (3), the longitudinal acceleration wave has catched up with the dislocation, while the transverse one still lags behind. The latter overcomes the dislocation only in pictures (4).

It is interesting to observe the reinforcement of the fields, on the part of the boundary of the longitudinal acceleration ring comprised between the two Mach cones, in components $\sigma_{xx}$ and $\sigma_{yy}$. These high-field segments have signs opposite to those of the longitudinal Mach cone, so that this region is subjected to a high stress gradient. In the example displayed, where the dislocation velocity is rather close to $\cL$, the longitudinal acceleration ring and braking ring closely follow each other, inducing a particularly strong effect observable in pictures (3) of the figures, near to the forward longitudinal wave front.

The $\sigma_{xy}$ component in Fig.\ \ref{fig:fig5} presents another interesting geometric effect: as the stress is negative on the slip plane ahead of the transverse acceleration ring [picture (1)], the latter screens out part of the right positive lobe of the new static field once it catches up with the pinned dislocation [picture (4)], and thus continues to play a non-negligible role long after the initial acceleration has taken place. Obviously, in all cases the full static fields displayed in Fig.\ \ref{fig:fig2} are retrieved in stable form only after the slowest (shear) initial acceleration wave has overcome the dislocation.

\section{Concluding discussion}
\label{sec7}
To summarize, we proposed in the field-theoretical framework of continuum dislocation theory a new approximate procedure to compute analytically fields radiated by dislocations undergoing non-uniform motion at arbitrary velocities ---including supersonic ones--- and along arbitrary paths. Our results hold for an unbounded, isotropic, linear-elastic medium. The procedure becomes exact for motions with piecewise-constant velocity function $\bV(t)$. Overall, our work hinges on technical results of two sorts:

1) First, after having clarified the fundamental distributional nature of the Green tensor, a so-called \emph{isotropic regularization} procedure has been employed to regularize fields produced by point sources. Using distributions, the theory of the non-uniform motion of straight dislocations becomes formally simpler: prior to regularization, all the terms that enter the integral solution of the Volterra problem [Eqs.\ (\ref{eq:voltintscrew}) and (\ref{eq:voltintedge})], are regular distributions and pseudofunctions of clear mathematical meaning, unlike in classical approaches where such terms are usually non-integrable. The bottom line is that with the help of these objects, the dynamic Volterra-dislocation theory becomes well-defined, i.e., free of so-called \emph{non-integrable singularities}. It then became possible to carry out all differentiations in the elastodynamic fields, and to handle the `non-integrable singularities' in a suitable mathematical way. In particular, the framework legitimates operations such as the interchange of integration and differentiation, which are ill-defined in the standard approach \citep{XM83}. Thus, the theory of dislocations in particular and, more generally, that of defects in the elastic continuum, has the theory of distributions as its natural background, as emphasized, e.g., by \cite{Pellegrini11}. However, only few authors have worked along this line. For static dislocations, \citet{Kunin,deWit2,deWit4} and \citet{Mura} used already some results of distribution theory. Of course, the statics of dislocations is much simpler than their dynamics.

We showed that the regularization could be implemented from the outset, i.e., at the most fundamental level of the elastodynamic Green's tensor and its gradient, by carrying out their spatial convolution with a specific isotropic $\delta$-sequence, used with a fixed width parameter representing the dislocation core width. The nicety is that in practice, this amounts to considering the analytic continuation to complex values of the time variable of the function associated with the distributional Green's tensor, as shown by Eqs.\ (\ref{eq:gisoijzz}) and (\ref{eq:Gisoijk}), in which the imaginary part of the time is proportional to the source width, divided by the relevant wavespeed. Thus, the isotropic regularization is very easy to implement in the isotropic case where the wavespeeds have closed-form analytical expressions. On the other hand, the same `trick' would need to be modified for anisotropic media for which (except in a few particular cases) the wavespeeds must in general be computed numerically \citep{AKIR09,Bacon}. It should be noted that the particular Somigliana dislocation of \cite{ESHE49} can as well be brought down to an analytic continuation of field expressions with respect to the space coordinate in the direction of motion instead of time \citep{Pellegrini11}. The latter regularization is by essence anisotropic, and might be better suited to anisotropic media. However, it was not employed here in view of the next point below, for which using the isotropic regularization proved a little easier. Whatever the exact approach employed, this shows that a powerful way of regularizing distributional Green's functions is to seek regularizations in the form of analytic continuations. The main virtue of such regularizations is to suppress the need for tracking wavefronts in subsequent calculations, so that the results can be applied without modifications to supersonic sources. More generally, it has been observed that techniques of analytic continuation greatly simplify the formulas involved in problems of moving dislocations \citep{Pellegrini11,Pellegrini12,PELL14}.

We expect the same method to apply as well to more complex Green's functions such as the one \citep{EATW82} adequate to problems with layered media or free surfaces \citep{STRO70,FREU73}, thus alleviating the need to consider separately the subsonic and supersonic cases as in traditional methods of solution. Moreover, provided that the Green's function is known in closed form this analytic-continuation approach should straightforwardly extend to coupled-physics problems, e.g., thermoelasticity \citep{BROC97}, which we must however leave to further work.

2) The above regularization step does not by itself produce Mach cones, since Green's functions can only generate circular wavefronts at each instant. To arrive at Mach cones, which are are caustics of circular wavefronts, we need a second type of results. The fields emitted by a moving dislocation involve convolution integrals over past times of expressions built from the regularized Green's function. We faced the problem of their numerical computation. To handle arbitrary dislocation motion, these time integrals have been split into secondary integrals over a discrete set of time intervals in which the dislocation velocity can be assumed constant. The latter assumption makes it possible to get those secondary integrals in closed form, thereby giving the dynamic fields in terms of time-discretized but closed-form expressions. Time integration provides expressions able to generate Mach cones. By this means, full-field dynamic maps of the stress field could be produced, even in instances of supersonic motion, which has not been previously done from analytical expressions, to our knowledge.

As we carried it out, this second step is much more specific to the isotropic problem at hand than the first one above. The key closed-form integrals of Appendix \ref{sec:appA} were first reported in \cite{PL2014}, where they were simply proved by differentiation ---few details being given as to their method of obtention. Suffice it to say that the method rests on representing two-dimensional vectors in the plane as complex numbers, which eases the integration over time to obtain indefinite integrals in terms of elementary functions, in full tensor form. From them stem the expressions for finite-time intervals used in the present paper. Their complexity is a consequence of the vector character of the velocity, which can take on any direction. It is very difficult to retrieve from them the known analytical expressions for steady subsonic motion, and this step is best done numerically from Eqs.\ (\ref{eq:BvVoltSteady}). This drawback is a relative one, if one bears in mind that those powerful expressions are able to account for regularized fields, Mach cones for both shear and longitudinal waves, and arbitrary velocity direction. It is not clear that like integrals could be arrived at in generalized problems involving free surfaces, layered media, or even anisotropic media. Should closed-form time integration prove unfeasible, numerical integration could be attempted, in the hope of benefitting from the smooth character of the regularized Green kernels. However, some difficulties might occur in the rendering of Mach cones. This would be worth investigating in the future.

Turning now to the physical content of the results, it must be emphasized that we restricted ourselves, for simplicity, to a rigid dislocation core size. Thus `relativistic' effects of dynamic core-width variations \citep{Pellegrini12,PELL14}, and their associated radiative contributions, are not accounted for. However, this should not be considered a limitation of the method. How to bypass this restriction, which is necessary to couple the present calculations to an equation of motion for dislocations, will be examined elsewhere in connection with the use of Eshelby's regularization. Indeed, although easier to implement, the isotropic regularization is ill-suited to handling Lorentz-contraction effects (the source must contract in the direction of motion only).

Obviously, the formalism does not need any modification to address dynamic nucleation or annihilation processes in the bulk of the material. By conservation of the dislocation density, such events involve \emph{pairs} of dislocations of opposite signs.\footnote{Such dipoles are two-dimensional counterparts of dislocation loops in three dimensions.} To account, e.g., for a nucleation event, one only needs to add the fields of each dislocation of the expanding pair, as computed via Eqs.\ (\ref{eq:iso3}). These fields mutually cancel out in the incipient state of pair nucleation when both dislocations are at rest with coinciding positions.

Finally, it should be remarked that the fields patterns in Sec.\ \ref{sec6} are (unsurprisingly) found symmetric, up to sign changes, on both sides of the glide plane. However, recent numerical work with a field model of continuum mechanics \citep{ZHAN15} suggests that non-linear elasticity might be responsible for a strong asymmetry of the fields, and in particular of the Mach cones where fields are strongest. Indeed, the latter work features asymmetric field patterns much alike those in some atomistic simulations \citep{LISH02,TSUZ09}. The ones by \cite{LISH02} concern tungsten ---an almost isotropic metal; hence, the main cause of asymmetry cannot reside in elastic anisotropy. Therefore, another conclusion of the present work is that such effects cannot be captured by linear elasticity alone.

Consequently, although the present dynamic fields expressions based on linear isotropic elasticity are quite specific to the problem at hand, and might suffer from some limitations from the physical standpoint, they are  appropriate to first investigations of collective radiative properties of dislocations ensembles by means of simulations of the type discussed in \citep{Gurr14}, in the scarcely explored high-velocity/high-acceleration range.

\section*{Acknowledgement}
M.L. gratefully acknowledges the grants obtained from the
Deutsche Forschungsgemeinschaft
(Grant Nos. La1974/2-1, La1974/2-2, La1974/3-1). Y.-P.\ P.\ gratefully thanks C.\ Denoual for having freely provided field maps from an independent dynamic phase-field code \citep{DENO04}, which greatly helped debugging the Fortran code developped for the numerical calculations.
\appendix
\section{Equal-time limits}
\label{sec:etlims}
\setcounter{equation}{0}
\renewcommand{\theequation}{\thesection.\arabic{equation}}
We demonstrate here the equal-time limit identities obeyed by the Green tensor, in the two-dimensional framework of the rest of the paper.
For the anti-plane strain case, these identities are well-known from the study of the Helmholtz equation, and read (see \cite{Barton}, p.\ 241)
\begin{align}
\lim_{t\to 0^+} G^+_{zz}(\rr,t)=0\,,\qquad \lim_{t\to 0}\partial_t G^+_{zz}(\rr,t)=\rho^{-1}\delta(\rr)\,.
\end{align}
We focus hereafter on the plane-strain case, using elements from Section \ref{sec4}. The strategy consists in first proving the equal-time identities (\ref{eq:equal-time-iso})
on the isotropic-regularized Green tensor, and then letting the regularizing size $\varepsilon$ go to zero, to retrieve those identities for $G^+_{ij}(\br,t)$, following the general principle expressed by Eqs.\ (\ref{eq:limitdistr}).

Starting from definition (\ref{eq:gisoij}), and introducing the tensor
\begin{align}
T_{ij}=\delta_{ij}-2 \frac{x_i x_j}{r^2}\,,
\end{align}
we cast the regularized Green tensor in the form \citep{PL2014}
\begin{align}
\label{eq:greencompact}
G_{ij}^{\rm iso}(\rr,t)&=\frac{\theta(t)}{4\pi\rho}\Re\sum_{p={\rm T},{\rm L}}\frac{1}{c_p}
\left\{\delta_{ij}\pm\frac{1}{r^2}[2(c_p t+\ii\varepsilon)^2-r^2]T_{ij}\right\}\frac{1}{\sqrt{(c_p t+\ii\varepsilon)^2-r^2}}\,,
\end{align}
where the sum is over the wavespeed index, and where the `plus' and `minus' signs apply to the transverse, and longitudinal terms,
respectively. Then, immediately,
\begin{align}
\label{eq:lim1}
\lim_{t\to 0^+}G_{ij}^{\rm iso}(\rr,0)&=-\frac{1}{4\pi\rho}\Re\sum_{p={\rm T},{\rm L}}\frac{\ii}{c_p}
\left\{\delta_{ij}\mp\frac{1}{r^2}[2\varepsilon^2+r^2]T_{ij}\right\}\frac{1}{\sqrt{r^2+\varepsilon^2}}=0\,,
\end{align}
since the expression under the $\Re$ operator is purely imaginary.

We next appeal to the following identities, demonstrated in Appendix A of \citep{PL2014}, where the Dirac terms originate from the branch cut of the complex square root function:
\begin{subequations}
\label{eq:idents}
\begin{align}
\frac{\partial}{\partial t}\sqrt{(c_p t+\ii\varepsilon)^2-r^2}
&=\frac{c_p(c_p t+\ii\varepsilon)}{\sqrt{(c_p t+\ii\varepsilon)^2-r^2}}+2\ii\sqrt{r^2+\varepsilon^2}\delta(t),\\
\frac{\partial}{\partial t}\frac{1}{\sqrt{(c_p t+\ii\varepsilon)^2-r^2}}
&=-\frac{c_p(c_p t+\ii\varepsilon)}{[(c_p t+\ii\varepsilon)^2-r^2]^{3/2}}-\frac{2\ii}{\sqrt{r^2+\varepsilon^2}}\delta(t)\,.
\end{align}
\end{subequations}
From these, we compute
\begin{align}
\partial_t G_{ij}^{\rm iso}(\rr,t)
&=\frac{\theta(t)}{4\pi\rho}\Re\sum_{p={\rm T},{\rm L}}\frac{c_p t+\ii\varepsilon}{[(c_p t+\ii\varepsilon)^2-r^2]^{3/2}}
\left\{-\delta_{ij}\pm\frac{1}{r^2}[2(c_p t+\ii\varepsilon)^2-3 r^2]T_{ij}\right\}\,,
\end{align}
in which the imaginary terms proportional to $\delta(t)$ in Eqs.\ (\ref{eq:idents}) have not survived due to the $\Re$ operator. Going to the limit $t\to 0$, we deduce
\begin{align}
\lim_{t\to 0^+}\partial_t G_{ij}^{\rm iso}(\rr,t)
&=\frac{1}{4\pi\rho}\Re\sum_{p={\rm T},{\rm L}}\frac{\varepsilon}{(r^2+\varepsilon^2)^{3/2}}
\left[\delta_{ij}\mp\frac{1}{r^2}(r^2+2\varepsilon^2)T_{ij}\right]=\rho^{-1}\frac{\delta_{ij}}{2\pi}\frac{\varepsilon}{(r^2+\varepsilon^2)^{3/2}}\,,
\end{align}
that is, by definition (\ref{reg-f}) of $\delta^\varepsilon(\rr)$,
\begin{align}
\label{eq:lim2}
\lim_{t\to 0^+}\partial_t G_{ij}^{\rm iso}(\rr,t)&=\rho^{-1}\delta_{ij}\delta^\varepsilon(\rr)\,.
\end{align}
Letting finally $\varepsilon\to 0^+$ in Eqs.\ (\ref{eq:lim1}) and (\ref{eq:lim2}) proves Eqs.\ (\ref{G-et-lim}).

\section{The upper boundary in time integrals}
\label{sec:tminus}
\setcounter{equation}{0}
The considerations put forward in Section~\ref{sec:remarks} are justified here, using $\Bv(\br,t)$ as an example;  one would proceed in the same manner with $\Bbeta(\rr,t)$. Let $\eta>0$ an infinitesimal number, and $t^\pm=t\pm\eta$. Consider first the traditional writing of the time-integral with $t^+$ as an upper boundary. From (\ref{v-M}), one reads
\begin{align}
v_i(\rr,t)&=\int_{-\infty}^{t^+}\d t'\int C_{jklm} G^+_{ij}(\rr-\rr', t-t') I_{lm,k}(\rr',t')\, \d \rr'\,.
\end{align}
Then,
\begin{align}
\partial_t v_i(\rr,t)&= C_{jklm}\left[\int G^+_{ij}(\rr-\rr',-\eta) I_{lm,k}(\rr',t^+)\d \rr'+\int_{-\infty}^{t^+}\hspace{-0.5em}\d t'\int \partial_t G^+_{ij}(\rr-\rr', t-t')I_{lm,k}(\rr',t')\d \rr'\right]\,.
\end{align}
The first term within brackets vanishes by the causality property (\ref{G-ret}). The second time-derivative reads
\begin{align}
\partial_t^2 v_i(\rr,t)&=C_{jklm}\left[\int\partial_t G^+_{ij}(\rr-\rr',-\eta)I_{lm,k}(\rr',t^+)\,\d \rr'
+\int_{-\infty}^{t^+}\hspace{-0.5em}\d t'\int\partial_t^2 G^+_{ij}(\rr-\rr',t-t')I_{lm,k}(\rr',t')\, \d \rr'\right]\nonumber\\
&=\int_{-\infty}^{t^+}\hspace{-0.5em}\d t'\int\partial_t^2 G^+_{ij}(\rr-\rr',t-t')C_{jklm}I_{lm,k}(\rr',t')\, \d \rr'
\end{align}
for the same reason. It follows that, by definition of the Green tensor,
\begin{align}
L_{ip} v_p(\rr,t)
&=\int_{-\infty}^{t^+}\d t'\int L_{ip}G^+_{pj}(\rr-\rr',t-t')C_{jklm}I_{lm,k}(\rr',t')\, \d \rr'\nonumber\\
&=\int_{-\infty}^{t+\eta}\d t'\int\delta(\rr-\rr')\delta(t-t')C_{iklm}I_{lm,k}(\rr',t')\, \d \rr'
=C_{iklm} I_{lm,k}(\rr,t')\,,
\end{align}
so that the integral solution indeed verifies the field equations of motion. This standard demonstration (e.g., \cite{Barton}) relies on causality and on the fact that $t^+>t$.

Consider now using $t^-$ as an upper boundary. Then,
\begin{align}
\partial_t v_i(\rr,t)&= C_{jklm}\left[\int G^+_{ij}(\rr-\rr',\eta)\, I_{lm,k}(\rr',t^-)\, \d \rr'+\int_{-\infty}^{t^-}\d t'\int \partial_t G^+_{ij}(\rr-\rr', t-t')\, I_{lm,k}(\rr',t') \d \rr'\right]\,,
\end{align}
so that
\begin{align}
\partial_t^2 v_i(\rr,t)&=
C_{jklm}\int G^+_{ij}(\rr-\rr',\eta)\dot{I}_{lm,k}(\rr',t^-)\, \d \rr'
+C_{jklm}\int\partial_t G^+_{ij}(\rr-\rr',\eta)\, I_{lm,k}(\rr',t^-)\,\d \rr'\nonumber\\
&{}+\int_{-\infty}^{t^-}\d t'\int\partial_t^2 G^+_{ij}(\rr-\rr',t-t')\, I_{lm,k}(\rr',t')\, \d \rr'\,.
\end{align}
By the equal-time limits (\ref{G-et-lim}), the first term in the left-hand side vanishes while the second one reduces to
\begin{align}
C_{jklm}\int\partial_t G^+_{ij}(\rr-\rr',\eta)\, I_{lm,k}(\rr',t^-)&=\rho^{-1}C_{iklm}I_{lm,k}(\rr,t)\,.
\end{align}
It follows that
\begin{align}
L_{ip} v_p(\rr,t)&=C_{iklm}I_{lm,k}(\rr,t)+\int_{-\infty}^{t-\eta}\d t'\int\delta(\rr-\rr')\delta(t-t')C_{iklm}I_{lm,k}(\rr',t')\, \d \rr'
=C_{iklm}I_{lm,k}(\rr,t)\,,
\end{align}
since now the interval of integration does not contain $t'=t$ any more. This second method thus relies on the equal-time limits rather than on causality. However, the same result is obtained in both cases, so that both writings of the integral are correct.


\section{The indefinite integral $J^{\rm iso}_{ijk}$}
\label{sec:appA}
\setcounter{equation}{0}
For brevity, the reference \citep{PL2014} is denoted as (PL) hereafter.
\subsection{Preliminary remarks and notations}
Although they are of a wider range of application, as the present work
demonstrates, the expressions that were given in (PL) for the quantity herein denoted by $J^{\rm iso}_{ijk}(\rr,t;\BV)$ were presented there in a way adapted to dislocations instantaneously accelerated from rest to constant velocity. The equations of immediate interest to us being somewhat scattered through the text of
the latter reference, the purpose of this Appendix is to summarize them neatly. The following intermediate quantities are employed:
\begin{subequations}
\label{eq:intermediate1}
\begin{align}
\bbeta&=\bV/c\,,\\
\bhn&=\BV/V=\Bbeta/\beta\,,\qquad \text{(velocity director)}\\
\bhr&=\rr/r\,,\\
\label{eq:calRdef}
{\boldsymbol{\cal{R}}}(\tau)&=\rr+\bV \tau\,,\\
\label{eq:Sdef}
S(\tau)&=\sqrt{c^2\tau^2-{\cal{R}}(\tau)^2}\,,\\
A^\pm_{ij}&=(1-\beta^2)(\delta_{ij}-\hn_i\hn_j)\pm\hn_i\hn_j\,,\\
X_{ij}&=r_i\beta_j-\beta_i r_j\,,
\end{align}
\end{subequations}
where $c$ is a generic placeholder for wavespeeds $\cT$ or $\cL$. When $V=0$, the unit director $\mathbf{\hat n}$ is arbitrary. Except for the index $z$, all indices below take on values $1$ or $2$. The introduction of the cross-product $X_{ij}$ allows for writings of expressions (\ref{eq:calJzzk}) and (\ref{eq:calJijk}) below shorter than the ones reported in (PL).

\subsection{Explicit expressions}
\subsubsection{Antiplane-strain case}
The expression of the only nonzero component of $J^{\rm iso}_{ijk}(\rr,t;\BV)$
is read from Eqs.\ (46--48) of (PL). Let
\begin{align}
\label{eq:calJzzk}
\mathcal{J}_{zzk}(\rr,\tau;\BV,c)&=S^{-1}\big(\br\cdot\mathsf{A}^+\cdot\br\big)^{-1}(X_{kl}{\cal{R}}_l-c\tau\,r_k)\,.
\end{align}
Introducing $\beta_{\rm T}=\bV/\cT$, we have (PL),
\begin{align}
\label{eq:Jisozzk}
J^{\rm iso}_{zzk}(\rr,t;\BV)&=\frac{1}{2\pi\mu}\Re\mathcal{J}_{zzk}(\rr-\ii\,\varepsilon \bbeta_{\text{T}},t+\ii\,\varepsilon/\cT;\BV,\cT)\,.
\end{align}
Thus, the function $\mathcal{J}_{zzk}(\rr,\tau;\BV,c)$ is used quite generally with complex-valued arguments $\br$ and $\tau$.

\subsubsection{Plane-strain case}
The following intermediate quantities are needed:
\begin{subequations}
\begin{align}
B_{mlx}&=2\hn_m\hn_l-\delta_{ml}\,,\qquad
B_{mly}=\epsilon_{zmp}\hn_p\hn_l+\epsilon_{zlp}\hn_p\hn_m\,,\\
Q_i&=S^{-1}{\cal{R}}_i\,,\\
q_i&=(\rr\cdot\mathsf{A}^+\cdot\rr)^{-1}r_i\,,\\
U_{ij}&=\delta_{ij}+Q_i Q_j,\qquad V_{ij}=(\rr\cdot\mathsf{A}^+\cdot\rr)^{-1}A^-_{ij},\qquad W_{ij}=\delta_{ij}-2\,q_m A^+_{mi}r_j\,,\\
\label{eq:Lxi}
L_{x,i}&=c\tau S^{-1}{\cal{R}}^{-2}{\cal{R}}_i\,,\\
L_{y,i}&=-\epsilon_{zip}\left[c\tau S^{-1}{\cal{R}}^{-2}{\cal{R}}_p+(\bbeta\cdot\mathbf{Q}-c\tau S^{-1})q_p\right]\,,\\
\label{eq:Lxik}
L_{x,ik}&=c\tau S^{-1} {\cal{R}}^{-2}\left(U_{ik}-2 {\cal{R}}^{-2}{\cal{R}}_i
{\cal{R}}_k\right)\,,\\
L_{y,ik}&=-\epsilon_{zip}\left[L_{x,pk}+S^{-1}\left(U_{kl}\beta_l-c\tau S^{-1}Q_k\right)q_p
+\left(\bbeta\cdot\mathbf{Q}-c\tau S ^{-1}\right)(\br\cdot\mathsf{A}^+\cdot\br)^{-1}W_{kp}\right]\,.
\end{align}
\end{subequations}
The result to be given is built from the third-rank tensor [Eq.\ (89) in (PL)]
\begin{align}
\mathcal{J}_{ijk}(\br,\tau;\BV&,c)=
\left(2\beta^2\right)^{-1}
\bigl\{-S^{-1}(U_{kj}\beta_i+U_{ki}\beta_j)
+\left[(W_{ki}\beta_l-\beta_i W_{kl})V_{mj}+(W_{kj}\beta_l-\beta_j W_{kl})V_{mi}\right]r_m Q_l
\nonumber\\
\label{eq:calJijk}
&{}+\left(X_{il}V_{kj}+X_{jl}V_{ki}+2\,X_{kl}V_{ij}\right)Q_l+S^{-1}\left(V_{im}X_{jl}+V_{jm}X_{il}\right)r_m U_{kl}\nonumber\\
&{}-c\tau S^{-1}\left[S^{-1} Q_k\left(V_{im}r_j+V_{jm}r_i\right)r_m+\left(V_{ik}r_j+V_{jk}r_i+2 V_{ij}r_k\right)
+\left(V_{im}W_{kj}+V_{jm}W_{ki}\right)r_m\right]\nonumber\\
&{}-(L_{l,i}B_{klj}+L_{l,j}B_{kli}+2 L_{l,k} B_{ilj})
-r_m\left(L_{l,ik}B_{mlj}+L_{l,jk}B_{mli}\right)
\bigr\}\,.
\end{align}
By slightly modifying the above definitions of $Q_i$, $L_{j,i}$ and $L_{j,ik}$ in an obvious way, an overall factor $S^{-1}$ could be factored out in this formidable expression. This is not done here because we do not want to divert too much from the notations of (PL).

One immediately deduces from the equations presented in Section 4.4 of (PL)
that\footnote{Equation (\ref{eq:Jisoijk}) has no true counterpart in (PL), where $\smash{\mathcal{J}_{ijk}(\br,\tau;\BV,c)}$ was denoted as $\smash{J_{ijk}(\rr,\tau)}$, and where a quantity $\smash{I^{\rm iso}_{ijk}(\rr+\BV t,t)}$, equal to $\smash{J^{\text{iso}}_{ijk}(\rr,t;\BV)}-\smash{J^{\text{iso}}_{ijk}(\rr,0^+;\BV)}$ in the present notations, was introduced with a dependence in the $\rr$, $\BV$ and $t$ variables that acknowledges its co-moving nature. Our notational changes, which include the appearance of the generic wavespeed $c$ as an argument in $\mathcal{J}_{ijk}(\rr,\tau;\BV,c)$, make our equations easier to understand.}
\begin{align}
\label{eq:Jisoijk}
J^{\text{iso}}_{ijk}(\rr,t;\BV)&=J^{\text{iso}}_{zzk}(\rr,t;\BV)\delta_{ij}
+\frac{1}{2\pi\rho}\Re\sum_{P={\rm T},{\rm L}}\pm \frac{1}{c_P^2}\mathcal{J}_{ijk}(\br-\ii\varepsilon\bbeta_P,t+\ii\varepsilon/c_P;\BV,c_P)\,.
\end{align}
where $\bbeta_P=\BV/c_P$, $c_P$ takes on values $\cT$ or $\cL$, and where the `plus' and 'minus' signs apply, respectively, to the $\text{T}$-term and $\text{L}$-term in the sum. Longitudinal and transverse contributions are conspicuous. Again, the function $\mathcal{J}_{ijk}(\rr,\tau;\BV,c)$ is used in general with complex-valued arguments $\br$ and $\tau$.

\subsubsection{Remarks}
\label{sec:appAremarks}
First, it is emphasized that expressions (\ref{eq:Jisozzk}) and (\ref{eq:Jisoijk}) are written \emph{in the co-moving frame}. This is the reason why they feature the quantity ${\boldsymbol{\cal{R}}}$ introduced in Eq.\ (\ref{eq:calRdef}); see also remark following Eqs.\ (\ref{eq:isounif}).

Next, it should be pointed out that, as an indefinite integral over time, $J^{\rm iso}_{ijk}(\rr,t;\BV)$ is determined up to an arbitrary time-independent integration constant. This irrelevant term vanishes in the subtractions that define expressions (\ref{eq:igam2}) and (\ref{eq:iiso0J}), while $J^{\rm iso}_{ijk}(\rr,t;\BV)$ itself is of no definite physical significance.

Moreover, we observe that Eqs.\ (\ref{eq:Jisozzk}) and (\ref{eq:Jisoijk}) make use of the simultaneous regularizing substitutions $\rr\to \rr^{\text{shift}}=\rr-\ii\epsilon\Bbeta$ and $\tau\to\tau+\ii\epsilon/c$, where $\varepsilon$ is the finite core size, which leaves the vector ${\boldsymbol{\cal{R}}}$ invariant. For this, an analytic continuation to complex times and positions of the functions $\mathcal{J}_{ijk}(\rr,\tau;\BV,c)$ is required. Accordingly, all square roots are used in the sense of the principal determination ---the one usually implemented in computers. With this convention the following limits hold:
\begin{subequations}
\begin{align}
\lim_{r\to 0}r^{\text{shift}}&=\lim_{r\to 0}\sqrt{(\rr-\ii\varepsilon\Bbeta)\cdot(\rr-\ii\varepsilon\Bbeta)}=-\ii\sign(\bhr\cdot\bhn)\beta\varepsilon\,,\\
\lim_{r\to 0}\bhr^{\text{shift}}&=\lim_{r\to 0}\frac{\br^{\text{shift}}}{r^{\text{shift}}}=\sign(\bhr\cdot\bhn)\bhn\,.
\end{align}
\end{subequations}
Thus, the continuation of the norm $r=\sqrt{\br\cdot\br}$ of $\br$ is in general \emph{complex-valued} (not a norm any more).

Finally, we observe that by going to the limit $\varepsilon\to 0$, expressions (\ref{eq:Jisozzk}) and (\ref{eq:Jisoijk}) become distributions, with Mach cones in the form of Dirac-like measures for faster-than-wave velocities. Extracting this limit for arbitrary velocities is thus a complicated business in general, an instance of which has been given in (PL) in the antiplane-strain case. However, upon neglecting the distributional character of this limit, more classical expressions of the fields a Volterra dislocation are retrieved by letting $\varepsilon=0$ right away. Of course, such expressions are singular (i.e., infinite, hence not physically meaningful) at the dislocation position and at wavefronts, and hold only in the absence of Mach cones, i.e., for velocities less than $\cT$. Thus, for Volterra dislocations and $|V|<\cT$, we can take
\begin{subequations}
\label{eq:Volterra}
\begin{align}
J^{\rm iso}_{zzk}(\rr,t;\BV)|_{\varepsilon=0}&=\frac{1}{2\pi\mu}\Re\mathcal{J}_{zzk}(\rr,t;\BV,\cT)\,,\\
J^{\text{iso}}_{ijk}(\rr,t;\BV)|_{\varepsilon=0}&=J^{\text{iso}}_{zzk}(\rr,t;\BV)|_{\varepsilon=0}\,\delta_{ij}
+\frac{1}{2\pi\rho}\Re\sum_{P={\rm T},{\rm L}}\pm \frac{1}{c_P^2}\mathcal{J}_{ijk}(\br,t;\BV,c_P)\,.
\end{align}
\end{subequations}

\subsection{Particular values and limits}
The above expressions yield non-singular field values in particular limits of interest. Limiting forms are rather difficult to extract in the plane-strain case, due to the complexity of (\ref{eq:calJijk}). The expressions reported below have been checked with the help of a symbolic computational toolbox.

\subsubsection{Values at $\tau=0$}
\label{sec:valtzer}
For $\tau=0$, the quantity $S(\tau)$ becomes $S=\ii\,r$, while $Q_i=-\ii\,\hr_k$ and $U_{ij}=\delta_{ij}-\hr_i\hr_j$. It follows that in the particular case where $\br$ and $\tau$ are taken real-valued, $\mathcal{J}_{zzk}$ and $\mathcal{J}_{ijk}$ are purely imaginary. Hence,
\begin{align}
\lim_{\tau\to 0^+}\Re\mathcal{J}_{zzk}(\rr,\tau;\BV,c)&=0,\qquad
\lim_{\tau\to 0^+}\Re\mathcal{J}_{ijk}(\rr,\tau;\BV,c)=0\qquad\text{($\br$,$\tau$ $\in$ $\mathbb{R}$)}\,.
\end{align}
When $\varepsilon=0$, inserting these expressions into Eqs.\ (\ref{eq:Volterra})  we conclude  that
\begin{align}
J^{\text{iso}}_{zzk}(\rr,0^+;\BV)|_{\varepsilon=0}&=0,\qquad J^{\text{iso}}_{ijk}(\rr,0^+;\BV)|_{\varepsilon=0}=0\,.
\end{align}

\subsubsection{Limit $\tau\to+\infty$}
\label{sec:applimtauinf}
This limit is needed to compute the upper boundary term with $\tau=+\infty$ in expression (\ref{eq:iiso0J}). Letting $u=\bhr\cdot\bhn$, we introduce the notation
\begin{align}
\mathcal{A}=r^{-2}(\rr\cdot\mathsf{A}^+\cdot\rr)&=(1-\beta^2)+\beta^2 u^2\,.
\end{align}
A straightforward but very long calculation yields
\begin{subequations}
\begin{align}
\label{eq:limitauinfzzk}
\lim_{\tau\to\infty}\mathcal{J}_{zzk}(\rr,\tau;\BV,c)&=
-\frac{1}{r\sqrt{1-\beta^2}\mathcal{A}}\left[(1-\beta^2)\hr_k+\beta^2 u\,\hn_k\right]\,,\\
\label{eq:limitauinfijk}
\lim_{\tau\to\infty}\mathcal{J}_{ijk}(\rr,\tau;\BV,c)&=\frac{\sqrt{1-\beta^2}}{r\beta^2\mathcal{A}^2}\Bigl\{
-[(1-\beta^2)\mathcal{A}+2\beta^2\,u^2]\delta_{ij}\hr_k
+\beta^2\mathcal{A}(\hr_i\delta_{jk}+\delta_{ik}\hr_j)\nonumber\\
&\hspace{1cm}{}
-\beta^2(\mathcal{A}-2)u\,\delta_{ij}\hn_k
-(1+\beta^2)\mathcal{A}\,u(\hn_i\delta_{jk}+\hn_j\delta_{ik})\nonumber\\
&\hspace{1cm}{}
+[(1-\beta^2)\mathcal{A}+2\beta^4\,u^2](\hr_i\hn_j+\hn_i\hr_j)\hn_k\nonumber\\
&\hspace{1cm}{}
+[(2-\beta^2)\mathcal{A}+2\beta^4\,u^2]\hn_i\hn_j\hr_k\nonumber\\
&\hspace{1cm}{}
+\beta^2(1-\beta^2)^{-1}[(2-3\beta^2)\mathcal{A}+2\beta^4\,u^2]u\,\hn_i\hn_j\hn_k\nonumber\\
&\hspace{1cm}{}
+2\beta^2(1-\beta^2)\,u(\hn_i\hr_j+\hr_i\hn_j)\hr_k
-2\beta^4\,u\,\hr_i\hr_j\hn_k-2\beta^2(1-\beta^2)\hr_i\hr_j\hr_k
\Bigr\}\,.
\end{align}
\end{subequations}
Using these expressions in Eqs.\ (\ref{eq:Jisozzk}) and (\ref{eq:Jisoijk}), which involves using complex values of their argument $\br$, gives access to the quantities
\begin{align}
J^{\text{iso}}_{zzk}(\rr,+\infty;\BV)\qquad\text{and}\qquad J^{\text{iso}}_{ijk}(\rr,+\infty;\BV)\,.
\end{align}
The above expressions can also be inserted right away into Eqs.\ (\ref{eq:Volterra}) to compute, for Volterra dislocations and velocity $|V|<\cT$ the quantities $J^{\text{iso}}_{zzk}(\rr,+\infty;\BV)|_{\varepsilon=0}$ and $J^{\text{iso}}_{ijk}(\rr,+\infty;\BV)|_{\varepsilon=0}$.

\subsubsection{Limit $\BV\to\boldsymbol{0}$}
This limit is needed for static field expressions. We have (PL)
\begin{subequations}
\begin{align}
\label{eq:limiVzerzzk}
\lim_{V\to 0}\mathcal{J}_{zzk}(\rr,\tau;\BV,c)&=-\frac{c\tau}{r\sqrt{c^2\tau^2-r^2}}\hr_k\,,\\
\label{eq:limiVzerijk}
\lim_{V\to 0}\mathcal{J}_{ijk}(\rr,\tau;\BV,c)&=\frac{c\tau}{r\sqrt{c^2\tau^2-r^2}}\,\hr_i\hr_j\hr_k
-\frac{c\tau}{r^3}\sqrt{c^2\tau^2-r^2}\,T_{ijk}(\bhr)\,,
\end{align}
\end{subequations}
where the following third-rank tensor has been used:
\begin{align}
\label{eq:tijk}
T_{ijk}(\bhr)=\delta_{jk}\hr_i+\delta_{ik}\hr_j+\delta_{ij}\hr_k-4\,\hr_i\hr_j\hr_k\,.
\end{align}
Expansions to second order in powers of $\bbeta$ would be needed to retrieve  Eq.\ (\ref{eq:limiVzerijk}) from expression (\ref{eq:calJijk}), which is inversely proportional to $\beta^2$. A shorter route is discussed in (PL).

\subsubsection{Double limit $\tau\to\infty$, $\BV\to\boldsymbol{0}$}
\label{sec:doublelim}
From (\ref{eq:limitauinfzzk}) and (\ref{eq:limiVzerzzk}), it is clear that the limits commute for $\mathcal{J}_{zzk}(\rr,\tau;\BV,c)$: one finds
\begin{align}
\label{eq:limlimzzk}
\lim_{\tau\to\infty}\lim_{V\to 0}\mathcal{J}_{zzk}(\rr,\tau;\BV,c)&=\lim_{V\to 0}\lim_{\tau\to\infty}\mathcal{J}_{zzk}(\rr,\tau;\BV,c)=-\frac{1}{r}\hr_k\,.
\end{align}
The case of $\mathcal{J}_{ijk}(\rr,\tau;\BV,c)$ is more subtle. Carrying out a Laurent expansion of (\ref{eq:limitauinfijk}) near $V=0$ ($\beta=0$), one gets
\begin{align}
\lim_{\tau\to\infty}\mathcal{J}_{ijk}(\rr,\tau;\BV,&c)
=\frac{c^2}{r}\left[
-\delta_{ij}\hr_k
-u(\hn_i\delta_{jk}+\hn_j\delta_{ik})
+(\hr_i\hn_j+\hr_j\hn_i)\hn_k
+2\hn_i\hn_j\hr_k
\right]V^{-2}\nonumber\\
&+\frac{1}{r}
\biggl[
\frac{1}{2}(1-2 u^2)\delta_{ij}\hr_k
+(\hr_i\delta_{jk}+\hr_j\delta_{ik})
+u\,\delta_{ij}\hn_k
-\frac{1}{2}u(3-2u^2)(\hn_i\delta_{jk}+\hn_j\delta_{ik})\nonumber\\
&{}-\frac{1}{2}(1+2 u^2)(\hr_i\hn_j+\hr_j\hn_i)\hn_k
+2 u(\hr_i\hn_j+\hr_j\hn_i)\hr_k
-2 u^2 \hn_i\hn_j\hr_k
+2 u\,\hn_i\hn_j\hn_k
-2\hr_i\hr_j\hr_k
\biggr]\nonumber\\
+{\rm O}\left(V^2\right)\,.
\end{align}
Despite appearances, it can be shown that the ${\rm O}(V^0)$ term in this expansion does not depend on the director $\bhn$ (!). The task is most easily performed by introducing polar angles for $\bhr$ and $\bhn$, and reducing each tensor component of the term by means of trigonometric identities, possibly with the help of an algebraic computational toolbox. The result is equal to
\begin{subequations}
\begin{align}
\label{eq:limtauinfijkLaur}
\lim_{\tau\to\infty}\mathcal{J}_{ijk}(\rr,\tau;\BV,&c)
=\frac{c^2}{r}\left[
-\delta_{ij}\hr_k
-u(\hn_i\delta_{jk}+\hn_j\delta_{ik})
+(\hr_i\hn_j+\hr_j\hn_i)\hn_k
+2\hn_i\hn_j\hr_k
\right]V^{-2}\nonumber\\
&+\frac{1}{r}\left[\hr_i\hr_j\hr_k+\frac{1}{2}T_{ijk}(\bhr)\right]+{\rm O}\left(V^2\right)\,.
\end{align}
On the other hand, Laurent-expanding (\ref{eq:limiVzerijk}) near $\tau=\infty$ directly gives
\begin{align}
\label{eq:limVzerinfijkLaur}
\lim_{V\to 0}\mathcal{J}_{ijk}(\rr,\tau;\BV,c)&=-\frac{c^2\tau^2}{r^3}T_{ijk}(\bhr)+\frac{1}{r}\left[\hr_i\hr_j\hr_k+\frac{1}{2}T_{ijk}(\bhr)\right]+{\rm O}\left(\tau^{-2}\right)\,.
\end{align}
\end{subequations}
The leading terms in expansions (\ref{eq:limtauinfijkLaur}) and  (\ref{eq:limVzerinfijkLaur}) are different from one another, and blow-up in the limits considered. However, they are physically irrelevant and can be disregarded because they are proportional to $c^2$: such terms (not containing $c$ in any other way) are eliminated between the transverse and longitudinal parts in Eq.\ (\ref{eq:Jisoijk}). Since the finite next-to-leading-order terms coincide, we can consider in practice that the limits commute as well for $\mathcal{J}_{ijk}(\rr,\tau;\BV,c)$, as far as $J^{\text{iso}}_{ijk}$ is concerned.

\subsubsection{Limit ${\boldsymbol{\cal{R}}}\to\boldsymbol{0}$}
This limit corresponds to letting $\rr\to -\BV\tau$ in the co-moving frame, which provides the fields at the origin of coordinates in the laboratory frame. In principle, this point should be of no particular interest, except in the static case $V=0$ where it locates the center of the dislocation core. However, since, e.g., $L_{x,i}$ in (\ref{eq:Lxi}) has one term proportional to $1/\mathcal{R}$ and moreover $L_{x,ik}$ in (\ref{eq:Lxik}) has one term proportional to $1/\mathcal{R}^2$, we need to verify that no ill-definiteness arises in this limit. Assuming that $V\not=0$ the following limit and Taylor expansion are obtained:
\begin{subequations}
\begin{align}
\label{eq:apslimRzer}
\lim_{\rr\to -\BV\tau}\mathcal{J}_{zzk}(\rr,\tau;\BV,c)&=\frac{1}{V\tau}\hn_k\,,\\
\label{eq:pslimRzer}
\mathcal{J}_{ijk}(\rr,\tau;\BV,c)&=\frac{c^2}{V^2\mathcal{R}}\left(1+\frac{V\tau}{\mathcal{R}}\right)T_{ijk}(\bhn)
+\frac{c^2-(V/2)^2}{V^3\tau}T_{ijk}(\bhn)-\frac{1}{4 V\tau}(\delta_{jk}\hn_i+\delta_{ik}\hn_j+\delta_{ij}\hn_k)\nonumber\\
&{}+\text{O}\bigl(\mathcal{R}\bigr)\,.
\end{align}
\end{subequations}
Thus, the antiplane-strain limit (\ref{eq:apslimRzer}) is not problematic, whereas the plane-strain expansion (\ref{eq:pslimRzer}) is singular as $\mathcal{R}\to 0$. However the divergent terms, proportional to $c^2$, are irrelevant for the same reason as in the previous Section, and can be disregarded. In practice, one can thus consider that
\begin{align}
\lim_{\rr\to -\BV\tau}\mathcal{J}_{ijk}(\rr,\tau;\BV,c)
&\simeq -\frac{1}{4V\tau}\left[T_{ijk}(\bhn)+(\delta_{jk}\hn_i+\delta_{ik}\hn_j+\delta_{ij}\hn_k)\right]\,,
\end{align}
which is finite.


\end{document}